\documentclass[12pt]{article}

\usepackage[dvips]{graphicx}
\usepackage{epsfig}
\usepackage{amsmath,amsfonts,amssymb,amsthm}
\usepackage{mathrsfs,mathtools}
\usepackage{verbatim}
\usepackage{psfrag}
\usepackage{bm}
\usepackage{bbm}
\usepackage[utf8]{inputenc}
\usepackage[square,comma,sort&compress,numbers]{natbib}
\usepackage[dvipsnames]{xcolor}
\usepackage{slashed}
\usepackage{upgreek}
\usepackage[normalem]{ulem}
\usepackage{enumitem}
\usepackage{float}
\usepackage[T1]{fontenc}
\usepackage{soul}
\usepackage{pgfplots}
\usepackage{extarrows}
\pgfplotsset{compat=1.17}
\usetikzlibrary{external,decorations.pathmorphing}
\usepackage{cases}
\usepackage{listings}

\lstdefinestyle{mystyle}{
    basicstyle=\ttfamily\footnotesize,
    breakatwhitespace=false,         
    breaklines=true,                 
    captionpos=b,                    
    keepspaces=true,                 
    numbers=left,                    
    numbersep=5pt,                  
    showspaces=false,                
    showstringspaces=false,
    showtabs=false,                  
    tabsize=2
}
\usepackage{lmodern}
\usepackage{tikz}

\usepackage[T1]{fontenc}

\tikzset{
  external/only named=true,
  thick/.style={line width=.5pt},
  approximation/.style={line width=1.2pt},
  numerics/.style={black, dotted, line width=.8pt},
  amplitude/.style={dashed},
  estimate/.style={dashed, line width=.8pt},
  normal plot/.style={line width=.8pt},
}

\tikzset{snake it/.style={decorate, decoration=snake}}

\usepackage[colorlinks=true,urlcolor=blue,anchorcolor=blue,citecolor=blue,filecolor=blue
,linkcolor=blue,menucolor=blue,linktocpage=true,pdfproducer=medialab,pdfa=true
]{hyperref}

\usepackage{epsf,epsfig}
\usepackage{graphics}
\usepackage{subcaption}

\newcommand\blfootnote[1]{%
  \begingroup
  \renewcommand\thefootnote{}\footnote{#1}%
  \addtocounter{footnote}{-1}%
  \endgroup
}

\def\vec{\mathbf}

\usepackage[errorstop]{feynmp}
\newcommand{\lsim}
{\;\raisebox{-.3em}{$\stackrel{\displaystyle <}{\sim}$}\;}
\newcommand{\gsim}
{\;\raisebox{-.3em}{$\stackrel{\displaystyle >}{\sim}$}\;}

\newcommand{\rom}[1]{\uppercase\expandafter{\romannumeral #1\relax}}

\begin{document}                                                                                                                                                                                                                                                                                                                                                                                                                                                                                                                                                                                                                                                                    

\thispagestyle{empty}

\begin{flushright}
{
\small
KCL-PH-TH/2023-19
}
\end{flushright}

\vspace{-0.5cm}

\begin{center}
\Large\bf\boldmath
Model-independent bubble wall velocities in local thermal equilibrium
\unboldmath
\end{center}

\vspace{-0.2cm}

\begin{center}
Wen-Yuan Ai,$^{*1}$\blfootnote{$^*$wenyuan.ai@kcl.ac.uk} Benoit Laurent$^{\dagger 2}$\blfootnote{$^\dagger$benoit.laurent@mail.mcgill.ca} and Jorinde van de Vis$^{\ddagger 3}$\blfootnote{$^\ddagger$j.m.vandevis@uu.nl} \\
\vskip0.4cm

{\it $^1$Theoretical Particle Physics and Cosmology, King’s College London,\\ Strand, London WC2R 2LS, United Kingdom} \\
{\it $^2$McGill University, Department of Physics, 3600 University St.,\\
Montr\'{e}al, QC H3A2T8 Canada} \\
{\it $^3$Institute for Theoretical Physics, Utrecht University,\\
Princetonplein 5, 3584 CC Utrecht, The Netherlands}
\vskip1.cm
\end{center}

\begin{abstract}
Accurately determining bubble wall velocities in first-order phase transitions is of great importance for the prediction of gravitational wave signals and the matter-antimatter asymmetry. However, it is a challenging task which typically depends on the underlying particle physics model. Recently, it has been shown that assuming local thermal equilibrium can provide a good approximation when calculating the bubble wall velocity. In this paper, we provide a model-independent determination of bubble wall velocities in local thermal equilibrium. Our results show that, under the reasonable assumption that the sound speeds in the plasma are approximately uniform, the hydrodynamics can be fully characterized by four quantities: the phase strength $\alpha_n$, the ratio of the enthalpies in the broken and symmetric phases, $\Psi_n$, and the sound speeds in both phases, $c_s$ and $c_b$. 
We provide a code snippet that allows for a determination of the wall velocity and energy fraction in local thermal equilibrium in any model.
In addition, we present a fit function for the wall velocity in the case $c_s = c_b = 1/\sqrt 3$.

\end{abstract}

\newpage

\hrule
\tableofcontents
\vskip .85cm
\hrule

\newpage

\section{Introduction}
\label{sec:Intro}

The possibility that the Universe  underwent one or multiple first-order phase transitions (FOPTs) is a very intriguing one. Not only could such a FOPT provide one of the necessary conditions for the generation of the matter-antimatter asymmetry~\cite{Kuzmin:1985mm,Morrissey:2012db,Garbrecht:2018mrp}, but such a process could also generate a stochastic background of gravitational waves (GWs)~\cite{Witten:1984rs,Kosowsky:1991ua,Kosowsky:1992vn,Kamionkowski:1993fg,Huber:2008hg,Hindmarsh:2013xza}, which could be observable with the next generation of GW detectors~\cite{Grojean:2006bp,Caprini:2018mtu,Caprini:2019egz,LISACosmologyWorkingGroup:2022jok}. 

Usually, a FOPT is induced by the dynamics of a scalar field, which can be either fundamental or composite, and has a potential with at least two non-degenerate minima. The scalar field represents the order parameter of the phase transition. 
The transition proceeds by nucleation and expansion of bubbles. 
Inside of the bubbles, the scalar field is in the energetically favored minimum of the potential, and outside in the metastable minimum. Having a  symmetry breaking phase transition
in mind, we shall call the phase in front of the bubble wall the symmetric phase and the phase behind the wall the broken phase. Typically, the bubble wall initially accelerates after formation, until it reaches a terminal velocity $\xi_w$, due to interaction with the plasma surrounding the bubble walls. The value of this terminal wall velocity is of great phenomenological interest, as it affects the value of the matter-antimatter asymmetry if the latter was formed during the phase transition \cite{Cline:2020jre, Cline:2021dkf,Ellis:2022lft}, and the shape and amplitude of the generated GW signals~\cite{Espinosa:2010hh,Caprini:2015zlo, Gowling:2021gcy}. 

Early estimates of the wall velocity were based on hydrodynamic arguments.
Originally it was argued in Ref.~\cite{Steinhardt:1981ct} that, like in chemical combustion, the wall velocity should be fixed by the Chapman-Jouguet condition which leads to only Jouguet detonations. It was however shown in~\cite{Laine:1993ey} that the Chapman-Jouguet condition is unrealistic for cosmological phase transitions and therefore more types of solutions are possible~\cite{Laine:1993ey,Kurki-Suonio:1995rrv}. To replace the Chapman-Jouguet condition, an effective friction term proportional to a phenomenological coefficient can be added to the equation of motion of the scalar field~\cite{Ignatius:1993qn,Heckler:1994uu,Kurki-Suonio:1996gkq,Espinosa:2010hh,Huber:2011aa,Huber:2013kj}. This term parameterizes the backreaction of the fluid onto the expanding bubbles. Estimates of the friction parameter appear in Refs.~\cite{Heckler:1994uu,Huber:2011aa,Huber:2013kj} and in Refs.~\cite{Dine:1992wr,Liu:1992tn,Moore:2000wx}, the friction on the bubble wall is directly determined using kinetic theory. 

Alternatively, the wall velocity can be determined by solving the scalar field equation of motion as well as the Boltzmann equations for the particles in the plasma. These plasma particles get pushed out of equilibrium by the passing wall, and the friction force is thus dominated by the particles which interact most strongly with the wall. This approach does not require an assumption for the friction parameter. Traditionally, the Boltzmann equations were solved by using  the so-called fluid Ansatz and taking moments~\cite{Moore:1995si,Moore:1995ua}. Recently, this approach was critically assessed by using alternatives to the conventional fluid Ansatz \cite{Laurent:2020gpg} and by taking more than the standard three moments \cite{Laurent:2022jrs}. Typically, the calculation of bubble wall velocities with out-of-equilibrium effects taken into account is model dependent. This model-dependence enters, e.g., via the collision terms.

Typically, a full computation of the bubble wall velocity including out-of-equilibrium effects is rather challenging, and has only been performed for a handful of models \cite{Liu:1992tn,Moore:1995si,Moore:1995ua,Dorsch:2018pat,Wang:2020zlf,Laurent:2022jrs,Jiang:2022btc}. In practice, the value of the wall velocity is often simply set to $\xi_w \rightarrow 1$, or treated as an unknown parameter. For some recent studies related to bubble growth, see Refs.~\cite{Hoche:2020ysm,Friedlander:2020tnq,Azatov:2020ufh,Cai:2020djd,Cline:2021iff,Bea:2021zsu,Bigazzi:2021ucw,Lewicki:2021pgr,Gouttenoire:2021kjv,Dorsch:2021nje,DeCurtis:2022hlx,Wang:2022txy,Lewicki:2022nba,Ai:2022kqm,GarciaGarcia:2022yqb,LiLi:2023dlc,Krajewski:2023clt}. 

Perhaps surprisingly, the plasma even exerts a pressure on the bubble wall when it is in local thermal equilibrium (LTE). In other words, the friction is not a purely out-of-equilibrium effect. The contribution to the backreaction force from equilibrium effects only was considered for the first time in Ref.~\cite{Ignatius:1993qn}. In Ref.~\cite{Konstandin:2010dm} it was demonstrated in a hydrodynamic description that heating of the plasma can obstruct the bubble expansion.  Recently, the effective friction and bubble wall velocity in LTE were further studied in Refs.~\cite{BarrosoMancha:2020fay,Balaji:2020yrx,Ai:2021kak}. 
Interestingly, it was noticed in Ref.~\cite{Ai:2021kak} that in LTE, due to the conservation of entropy, there is an additional matching equation in the hydrodynamic quantities near the bubble wall. This not only allows for a determination of the wall velocity in a {\it model-independent} way but also makes the computation extremely simple. One might worry that the condition of LTE is too ideal, and that the omission of out-of-equilibrium effects can not lead to a proper estimate of the wall velocity. However, it was recently shown in Ref.~\cite{Laurent:2022jrs}, in the singlet scalar extension of the Standard Model, that the contributions from out-of-equilibrium effects are typically subdominant, and LTE gives a reasonable estimate of the wall velocity.
Moreover, as out-of-equilibrium effects provide an additional source of friction, the results for LTE can be interpreted as an upper bound on the wall velocity. Yet, we will see that the LTE approximation is only reasonable for so-called deflagration and hybrid solutions, and including out-of-equilibrium effects is essential for the determination of the wall velocity of detonations.
 
In order to bridge the gap between the full model-dependent out-of-equilibrium computation and the simple estimate of $\xi_w \approx 1$, in this work we will use the LTE approximation to estimate the wall velocity. The analysis given in Ref.~\cite{Ai:2021kak} is restricted to the so-called bag equation of state, which describes the fluid as pure radiation in both phases, with a temperature-independent vacuum energy difference. In the present work, we generalize the analysis to more general equations of state, in addition to going beyond the planar wall limit taken in Ref.~\cite{Ai:2021kak}, and including so-called hybrid solutions to the hydrodynamic equations. We apply our analysis to the case when the sound speeds in the broken and symmetric phases are approximately temperature independent. It will be shown that the hydrodynamics in LTE are fully captured by four parameters: the phase strength $\alpha_n$, the ratio of the enthalpies in the broken and symmetric phases, $\Psi_n$, and the sound speeds in both phases, $c_s$ and $c_b$. The bubble wall velocity $\xi_w$, together with other quantities such as the kinetic energy fraction $K$ or the efficiency factor $\kappa$, can be determined from these four parameters. This makes our results widely applicable; as long as the speeds of sound are approximately constant, our results can be used as an estimate of the wall velocity and the kinetic energy fraction. 

The outline of the paper is as follows. In Sec.~\ref{sec:hydro}, we give a brief review of the hydrodynamics of expanding bubbles, and the corresponding matching equations. In Sec.~\ref{sec:model-ind-matching}, we introduce a model-independent approximation of the matching equations, and demonstrate how the equations depend on only a handful of thermodynamic quantities. The model-independent hydrodynamics are solved and the results are presented in Sec.~\ref{sec:tem-results}. We conclude in Sec.~\ref{sec:Conc}. Several technical details are collected in the Appendices. For the convenience of the reader, we have included a Python snippet in Appendix~\ref{ap:code}, which performs the computation of the wall velocity and efficiency factor in terms of a number of phase transition parameters.

\section{Hydrodynamic equations and matching conditions}\label{sec:hydro}

We are interested in the hydrodynamic equations describing the plasma that surrounds the expanding bubble. We describe the plasma as a perfect fluid with temperature $T$, characterized by a model-dependent pressure $p(\phi,T)$, which is set by the finite-temperature effective potential of the scalar field $\phi$,
\begin{equation}    
    p(\phi,T ) = -V_{\rm eff}(\phi,T)\,,
\end{equation}
and includes contributions from all (thermalized) degrees of freedom in the plasma. Our effective potential $V_{\rm eff}(\phi,T)$ includes also the $\phi$-independent term. In the symmetric phase $\phi =0$, and at any value of $T$, the value of $\phi$ in the broken phase can be determined by minimizing the potential. We can thus understand $p$ to be a function of the temperature only. The other thermodynamic quantities of interest are the energy density $e$, the enthalpy $\omega$ and the entropy $s$, which can be obtained from the pressure via
\begin{equation}
    e= T\frac{\partial p}{\partial T} - p\,, \qquad 
    \omega = T\frac{\partial p}{\partial T}\,, \qquad s = \frac{\omega}{T}\,.
\end{equation}
The energy-momentum tensor of the fluid is given by
\begin{align}\label{eq:emfluid}
    T^{\mu\nu}_f=(e+p)u^\mu u^\nu -(p+V(\phi))g^{\mu\nu}\,,
\end{align}
where $V(\phi)$ denotes the zero-temperature contribution to the pressure and $u^\mu$ is the fluid velocity. $g^{\mu\nu}$ denotes the spacetime metric, which we assume to be Minkowski. The fluid equations follow by projecting the continuity equations along the directions parallel and perpendicular to the fluid flow (see, e.g., Ref.~\cite{landau1987fluid})
\begin{subequations}
\label{eq:contituity}
\begin{align}
    &u_\nu \partial_\mu T^{\mu\nu}_f=0\,,\\
    &\Bar{u}_\nu\partial_\mu T_f^{\mu\nu}=0\,.
\end{align}
\end{subequations}
Here $\Bar{u}^\mu$ is the normalized vector orthogonal to $u^\mu$. Working in the frame where the center of the bubble is at rest, we have $u^\mu=\gamma(1,\vec{v})$ and $\Bar{u}^\mu=\gamma(v,\vec{v}/v)$, with $v = |\vec v|$ and the Lorentz factor $\gamma = 1/\sqrt{1-v^2}$. In a spherical coordinate system, $u^\mu=(\gamma,\gamma v,0,0)$ and $\Bar{u}^\mu=(\gamma v,\gamma,0,0)$. Since there is no characteristic scale in the problem, the solution is self-similar, i.e. it depends only on the dimensionless variable $\xi=r/t$, where $r$ is the radial distance from the center of the bubble and $t$ is the time since bubble nucleation.  The bubble wall velocity will be denoted by $\xi_w$.

Writing Eqs.~\eqref{eq:contituity} explicitly in spherical coordinates and using the relations between $\partial_t$, $\partial_r$ and $\partial_\xi$, one obtains 
\begin{subequations}
\begin{align}
   & (\xi-v)\frac{\partial_\xi e}{\omega}=2\frac{v}{\xi}+[1-\gamma^2v(\xi-v)]\partial_\xi v\,,\\
   & (1-v\xi)\frac{\partial_\xi p}{\omega}=\gamma^2(\xi-v)\partial_\xi v\,.
\end{align}
\end{subequations}
These two equations can be rearranged into 
\begin{subequations}
\label{eq:continuity2}
\begin{align}
    &2\frac{v}{\xi}=\gamma^2(1-v\xi)\left[\frac{\mu^2(\xi,v)}{c^2}-1\right]\partial_\xi v\,,\\
    &\partial_\xi\omega=\omega\left(1+\frac{1}{c^2}\right)\gamma^2\mu(\xi,v)\partial_\xi v\,,
\end{align}
\end{subequations}
where $c$ denotes the speed of sound
\begin{equation}
    c^2(T) = \frac{dp/dT}{de/dT}\,,
\end{equation}
and where the Lorentz-boosted velocity is given by
\begin{align}
    \mu(\xi,v)=\frac{\xi-v}{1-\xi v}\,.
\end{align}
Of particular interest is $\mu(\xi_w)$ which is the fluid velocity at the bubble wall viewed from the rest frame of the wall.  

From the solution of the hydrodynamic equations, the averaged kinetic energy density in the fluid is given by
\begin{equation}\label{eq:rhofl}
    \rho_{\rm fl}   = \frac{3}{\xi_w^3} \int d\xi\, \xi^2 v^2\gamma^2 \omega\,.
\end{equation}
From the above quantity, it is easy to obtain the kinetic energy fraction $K$ 
\begin{equation}
\label{eq:K-def}
    K = \frac{\rho_{\rm fl}}{e_n}\,,
\end{equation}
where the subscript $n$ denotes that the relevant quantity is to be evaluated in the symmetric phase and at the nucleation temperature. Here, we have followed the convention that the vacuum energy of the broken phase vanishes.  $K$ is one of the quantities that determine the amplitude of the GW signals from sound waves (see, e.g., Ref.~\cite{Caprini:2019egz}). In practice, it will be convenient to relate the kinetic energy fraction $K$ to the so-called efficiency factor $\kappa$,
 \begin{equation}
    \kappa = \frac{4\rho_{\rm fl}}{3 \alpha_n \omega_n}\,,
 \end{equation}
where $\alpha_n$ is the phase transition strength (see Eq.~(\ref{eq:defalpha}) and below). As demonstrated in Refs.~\cite{Giese:2020rtr,Giese:2020znk}, the efficiency factor can be determined from the hydrodynamic equations in a model-independent way, and eventually be converted into the kinetic energy fraction via $K =3\kappa \alpha_n \Gamma/4$, with $\Gamma = \omega_n/e_n$ being the adiabatic index. 

\begin{figure}[t]
    \centering
    \includegraphics[scale=0.3]{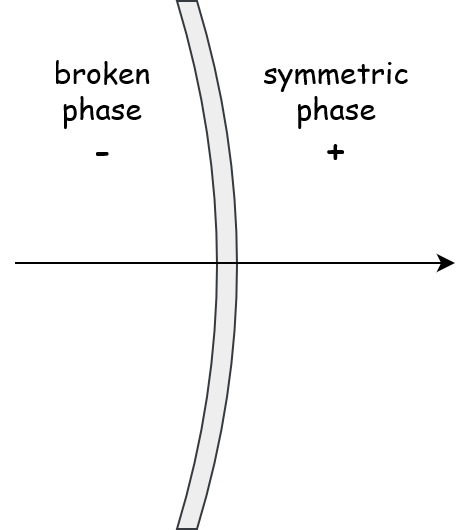}
    \caption{Rest frame of the bubble wall used for the matching conditions for hydrodynamic quantities in front of and behind the wall.}
    \label{fig:wall}
\end{figure}

Now we discuss the matching conditions for the hydrodynamic quantities in front of and behind the bubble wall. In this case, it is more convenient to work in the rest frame of the wall, see Fig.~\ref{fig:wall}. Integrating the condition of  energy-momentum conservation from just behind the bubble wall to just in front of the bubble wall, one obtains the following two well-known matching conditions
\begin{subequations}
\label{eq:junctionAB}
\begin{align}
    &\omega_+\gamma_+^2v_+=\omega_-\gamma_-^2v_-\, ,\label{eq:conditionA}\\
    &\omega_+\gamma_+^2v_+^2+p_+=\omega_-\gamma_-^2v_-^2+p_-\, ,\label{eq:conditionB}
\end{align}
\end{subequations}
where a subscript ``$\pm$'' is used to denote quantities in front of/behind the bubble wall. To be explicit, $\omega_+=\omega_s(T_+)$, $\omega_-=\omega_b(T_-)$ (and similarly for $p_{\pm }$), where the label ``$s/b$'' denotes symmetric/broken phase. Note that $v_+$, $v_-$ are the fluid velocities (defined to be positive) observed in the rest frame of the bubble wall. When comparing them to the fluid velocity in the rest frame where the center of the bubble is at rest, $v(\xi)$, one needs to take a Lorentz boost of $v(\xi)$, giving $\mu(\xi_w)$. 
Eqs.~\eqref{eq:conditionA} and~\eqref{eq:conditionB} can be rewritten as
\begin{align}\label{eq:matching}
    v_+v_-=\frac{p_+-p_-}{e_+ - e_-}\,,\quad \frac{v_+}{v_-}=\frac{e_- +p_+}{e_+ + p_-}\,.
\end{align}

The equations discussed above are well known and have been extensively studied in the literature. Given an equation of state, the values of $v_+, v_-, T_-$, and in principle, $T_+$ are unknown. However, for detonation solutions (as discussed below), $T_+$ is equal to the nucleation temperature $T_n$, and for deflagrations and hybrids, $T_+$ is also determined by $T_n$ and can be considered as given. This leaves us with three unknowns, which are subject to two equations. Typically, the degeneracy is resolved by fixing the value of the wall velocity $\xi_w$, either by explicit computation or estimation. The wall velocity sets either $v_+$ or $v_-$ (in the case of a detonation, $v_+ = \xi_w$; for a deflagration, $v_- = \xi_w$; and for a hybrid, $v_-$ equals the sound speed), and the two matching equations can then be used to determine $T_-$ and the other one in $\{v_+,v_-\}$.

In the present work, we will not impose the wall velocity, but instead solve it from the hydrodynamic equations. To this end, we consider an additional matching relation, due to entropy conservation:
\begin{equation} 
    s_+ \gamma_+ v_+ =s_- \gamma_- v_-\,.
\end{equation}
The above condition is a special situation of the non-negativity of entropy production~\cite{Laine:1993ey}. Using Eq.~\eqref{eq:conditionA}, it can be also written as
\begin{align}\label{eq:matchentropy}
T_+\gamma_+=T_-\gamma_-\,.
\end{align}
This is the matching condition first used in Ref.~\cite{Ai:2021kak} to determine the wall velocity. Since we now have three equations for the three unknowns $v_\pm, \,T_-$, we can determine all three quantities, and infer the value of the wall velocity, as pointed out in Ref.~\cite{Ai:2021kak}. There, the analysis was performed in the so-called bag equation of state, and we will now demonstrate how the results can be generalized to more general situations in a model-independent way. In Appendix~\ref{ap:LTEEntropy}, we demonstrate how the assumption of LTE in the scalar field equation of motion reduces the problem to a purely hydrodynamic one.

\section{Model-independent matching equations}
\label{sec:model-ind-matching}

Refs.~\cite{Giese:2020rtr,Giese:2020znk} demonstrated how the hydrodynamic equations can be solved in a model-independent manner in the case where only the two matching conditions Eq.~\eqref{eq:junctionAB} are used and the wall velocity is imposed. The purpose of Refs.~\cite{Giese:2020rtr,Giese:2020znk} was to express the kinetic energy fraction purely in terms of quantities that can be determined at the nucleation temperature and the wall velocity. 
To this end, a Python snippet was provided, which takes only the phase transition strength, the speed of sound, and the wall velocity as input parameters. These results can be seen as a generalization of Ref.~\cite{Espinosa:2010hh}, where the kinetic energy fraction was determined for the bag equation of state. In that case, the speed of sound is assumed to be $c^2 = 1/3$ and the kinetic energy fraction is a function of the phase transition strength and wall velocity only.
In this work, we will follow the approach developed in Refs.~\cite{Giese:2020rtr, Giese:2020znk}, and provide the wall velocity and the kinetic energy fraction as a function of the phase transition strength and the speed of sound in both phases. As we will see, an additional parameter, the ratio of enthalpies in both phases, is also required.

We first briefly summarize the approach of Refs.~\cite{Giese:2020rtr,Giese:2020znk} and demonstrate how it can be applied to the case with our new matching condition, which will allow for a model-independent determination of the wall velocity in LTE. 
Following Refs.~\cite{Giese:2020rtr,Giese:2020znk} we define the following notation:
\begin{align}
    \Delta X\equiv X_s(T_+)-X_b(T_-)&=\left[X_s(T_+)-X_b(T_+)\right]+\left[X_b(T_+)-X_b(T_-)\right]\notag\\
    &\equiv DX(T_+)+\delta X\,,
\end{align}
where $X$ denotes any thermodynamic quantity (e.g. $p, e, \omega$ or $s$). Here, the symbol $D$ denotes the difference of the quantities of the same type in the symmetric and broken phases evaluated at the same temperature which is not necessarily limited to $T_+$ and will be indicated explicitly. The matching equations for $v_\pm$, Eq.~\eqref{eq:matching}, can then be written as
\begin{equation}
\label{eq:matchingdelta}  
    \frac{v_+}{v_-} =\frac{1 - \Delta e/\omega_+}{1 - \Delta p/\omega_+}\,, \qquad v_+ v_- = \frac{\Delta p/\omega_+}{\Delta e/\omega_+}\,.
\end{equation}

At this point, it is particularly convenient to limit our discussion to the case where the sound velocities
are approximately constant in both phases. This approximation is reasonable when $c_{b,s}(T)$ depend only weakly on $T$, or when $T$ is approximately uniform, which implies $T_+\approx T_-\approx T_n$. The former case is expected to hold in most realistic scenarios where the Universe is radiation dominated, since we then have $c_{b,s}^2(T)\approx 1/3$. And the latter is obtained when the FOPT is not too strong. The sound velocities can then be evaluated with
\begin{align}
    c^2_{b,s}\approx c^2_{b,s}(T_n) = \left. \frac{dp_{b,s}/dT}{Td^2p_{b,s}/dT^2}\right\vert_{T_n}\,.
\end{align}
 The assumption of constant sound speed allows for a mapping onto a simple “template model'' (called the $\nu$-model in Ref.~\cite{Giese:2020rtr}),  which was first introduced in~\cite{Leitao:2014pda} and will be discussed in more detail in Sec.~\ref{sec:tem-results}. 
Using this approximation, and introducing the so-called pseudotrace $\bar \theta$, the corresponding phase transition strength $\alpha$ and the ratio of enthalpies $\Psi$, 
\begin{align}\label{eq:defalpha}
    \Bar{\theta}=\left(e-\frac{p}{c_b^2}\right)\,,\quad\alpha(T)=\frac{D\Bar{\theta} (T) }{3\omega_s(T)}\,,\quad \Psi(T)=\frac{\omega_b(T)}{\omega_s(T)}\,,
\end{align}
we show in Appendix \ref{ap:sound} that the matching equations \eqref{eq:matchentropy} and \eqref{eq:matchingdelta} reduce to 
\begin{subequations}
\label{eq:matchingEqs}
\begin{align}
    \label{eq:matchingEqsA}
    \frac{v_+}{v_-} &=\frac{v_+v_-(\nu-1)-1+3\alpha_+  }{v_+v_-(\nu-1)-1 +3v_+v_-\alpha_+}\, ,\\
    \label{eq:matchingEqsB}
    v_+v_- & = \frac{-\left(\frac{\gamma_+}{\gamma_-}\right)^{\nu} \Psi_+ + 1 - 3 \alpha_{+} }{3 \nu \alpha_{+}} \left[1-(\nu-1) v_+ v_-\right]\,,
\end{align}
\end{subequations}
where $\nu \equiv 1+1/c_b^2$, $\alpha_+ \equiv \alpha(T_+)$ and $\Psi_+\equiv \Psi(T_+)$.

For deflagration and hybrid solutions (to be discussed in Sec.~\ref{sec:tem-results}), $T_+$ is not known \textit{a priori}, which makes the calculation of $\alpha_+$ and $\Psi_+$ difficult. One must rely on the integration of Eqs.~\eqref{eq:continuity2} to compute $\omega_+$, which allows us to express $\alpha_+$ and $\Psi_+$ in terms of $\alpha_n \equiv \alpha(T_n)$ and $\Psi_n \equiv \Psi(T_n)$ as
\begin{subequations}\label{eq:TpFromTn}
\begin{align}
    \alpha_+ &= \frac{\mu-\nu}{3\mu}+\frac{\omega_n}{\omega_+}\left( \alpha_n -\frac{\mu-\nu}{3\mu} \right)\,, \\
    \Psi_+ &= \Psi_n\left(\frac{\omega_+}{\omega_n}\right)^{\nu/\mu-1}\,,
\end{align}
\end{subequations}
where $\omega_n\equiv \omega_s(T_n)$ and $\mu \equiv  1+1/c_s^2$. These relations hold in the limit of constant sound speed and can easily be confirmed with the equation of state introduced in Sec.~\ref{sec:tem-results}.

Another motivation for assuming the sound velocities to be constant is that they 
appear in the fluid equations \eqref{eq:continuity2}. It then becomes necessary to assume the sound speeds to be approximately temperature independent if we want to keep the discussion independent of the explicit expressions of the pressure and energy density. This approximation turns out to work very well in practice for typical models featuring a FOPT \cite{Giese:2020rtr,Giese:2020znk,Tenkanen:2022tly}. 

Inspection of Eqs.~\eqref{eq:continuity2}, \eqref{eq:matchingEqs} and \eqref{eq:TpFromTn} then reveals that all model-dependence in the hydrodynamic equations has been captured by the four parameters $\alpha_n$, $\Psi_n$, $c_b$ and $c_s$. They can all be computed from a particle physics model at the temperature $T_n$, which is assumed to be known beforehand.
%


\section{Template model and results}
\label{sec:tem-results}

For the numerical study, we work in the template model in which the sound speeds are exactly temperature independent. But the results can also be applied to other models when the sound speeds are only approximately temperature independent, see Appendix~\ref{ap:sound} for a discussion of the applicability of the approximation. The results presented in this section have been obtained with the Python code given in Appendix~\ref{ap:code}. 

The template model is a useful tool to find a particle-physics-model-independent solution to the hydrodynamic equations. It was introduced in
Ref.~\cite{Leitao:2014pda} as a generalization of the bag model, with the following equation of state 
\begin{align}
\label{eq:nu_eos}
    &e_s(T)=\frac{1}{3} a_+ (\mu-1) T^\mu+\epsilon\,,\qquad p_s(T)=\frac{1}{3}a_+ T^\mu-\epsilon\,,\\
    &e_b(T)=\frac{1}{3}a_- (\nu-1)T^\nu \,,\qquad\qquad\  p_b(T)=\frac{1}{3}a_- T^\nu\,.
\end{align}
Here, $\epsilon$ is temperature independent and parameterizes the vacuum energy and $\mu$, $\nu$ are constants related to the sound speed in the symmetric and broken phases 
through
\begin{align}
    \mu=1+\frac{1}{c^2_{s}}\,,\quad \nu=1+\frac{1}{c^2_{b}}\,.
\end{align}
The temperature-independent parameters $a_{\pm}$ are dimensionful. The bag model is recovered when $\mu=\nu=4$ and in that case, $a_{\pm}$ become dimensionless. This equation of state was used in Refs.~\cite{Giese:2020rtr,Giese:2020znk} to parameterize the kinetic energy fraction. 

For phase transitions that are dominated by the particle content of the Standard Model, one expects the speed of sound to remain rather close to $ c_{s,b} \approx 1/\sqrt 3$ \cite{Laine:2015kra,Tenkanen:2022tly}, and, depending on the required level of accuracy, computations in the bag model might suffice. For phase transitions taking place in a hidden sector, on the other hand, deviations from $c_{s,b} = 1/\sqrt 3$ can be significant \cite{Tenkanen:2022tly}. Especially in phase transitions in a strongly coupled sector, the sound speed can be very suppressed \cite{Bea:2021zsu,Ares:2020lbt}. Moreover, the value of $\Psi_n$ can also be significantly smaller than the typical value in Standard-Model-like phase transitions \cite{Janik:2022wsx}. It should be noted though, that the applicability of LTE in these set-ups has not been studied.

\paragraph{Detonations}

Supersonic walls ($v_+\geq c_{b}$) are described by detonation solutions. For these bubbles, no shock wave can propagate in front of the wall, which implies $v_+=\xi_w$, $T_+=T_n$, $\alpha_{+}=\alpha_{n}$ and $\Psi_+=\Psi_n$. To obtain a consistent solution, a rarefaction wave must be present behind the supersonic wall, satisfying the conditions $c_{b}\leq v_-\leq v_+$. From Eq.~\eqref{eq:matchingEqsA}, it can be seen that the upper bound is reached for $v_+=v_-=1$, while the lower bound is reached for $v_-=c_{b}$ and corresponds to a Jouguet detonation with a wall velocity
\begin{align}\label{eq:vJ}
    \xi_w=\xi_J=c_{b}\left(\frac{1+\sqrt{3\alpha_{n}(1-c^2_{b}+3c^2_{b} \alpha_{n})}}{1+3c^2_{b}\alpha_{n}}\right)\,.
\end{align}
The Jouguet velocity $\xi_J$ is the smallest wall velocity a physically consistent detonation solution is allowed to have.

More generally, Eq.~\eqref{eq:matchingEqsA} has the solution 
\begin{align}
    v_- = \frac{A+\sqrt{A^2-4c^2_{b} \xi_w^2}}{2\xi_w}\,,
\end{align}
with $A=\xi_w^2+c_{b}^2[1-3\alpha_{n}(1-\xi_w^2)]$. Substituting this result into Eq.~\eqref{eq:matchingEqsB}, one can finally solve for $\xi_w$. The detonation solution is independent of $c_s$.

\begin{figure}[t]
    \centering
    \includegraphics[width=0.6\linewidth]{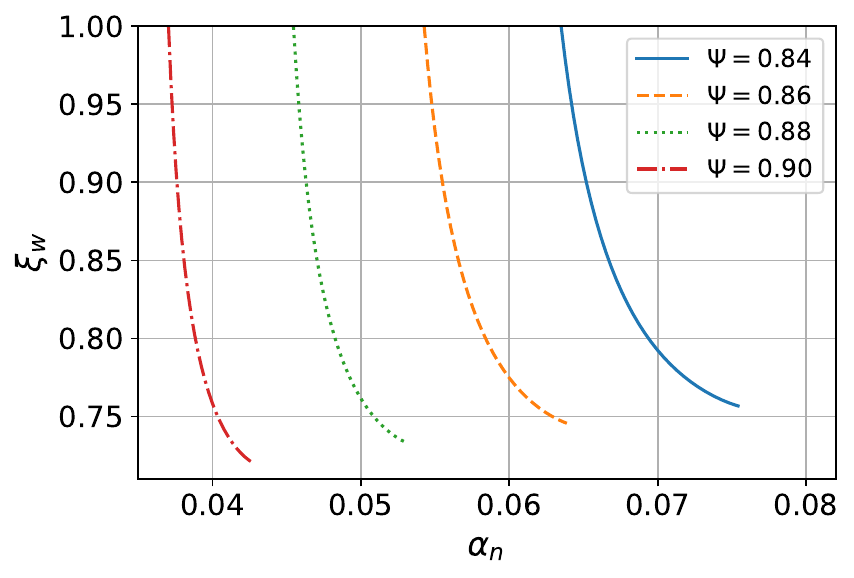}
    \caption{Wall velocity for detonation solutions as a function of $\alpha_n$ with $\nu=4$.}
    \label{fig:vw_det}
\end{figure}

Surprisingly, when solving all the matching equations as described above, we observe that the wall velocity is a decreasing function of $\alpha_{n}$, see Fig.~\ref{fig:vw_det}. This observation is consistent with Ref.~\cite{Ai:2021kak}.  Intuitively, this behavior can be explained by the fact that the friction slowing down the wall is maximized at the Jouguet velocity, when the hydrodynamic effects are the most violent. Thus, the fastest detonation solution ($\xi_w=1$) is reached when $\alpha_{n}$ is equal to its minimal value $\alpha_{n,\min}^{\rm det}$. When the deviation from the bag EOS is small ($\nu\approx 4$) and assuming $\alpha_n \ll 1$, one can show that 
\begin{align}
    \alpha_{n,\min}^{\rm det}\approx  \frac{1-\Psi_n}{12\Psi_n}[4-(1-\Psi_n )(\nu-4)]\,, \qquad {\rm for} \quad |\nu-4|\ll 1\,.
\end{align}
Note that when deriving the above result one cannot directly substitute the limit $v_+=v_-=1$ into Eq.~\eqref{eq:matchingEqsB} as the ratio $(\gamma_+/\gamma_-)$ is indeterminate in that limit.
On the other hand, the slowest detonation ($\xi_w=\xi_J$) is reached when $\alpha_{n}=\alpha_{n,\max}^{\rm det}$, which can be approximated when $\Psi_n \approx 1$ by
\begin{align}
    \alpha_{n,\max}^{\rm det}\approx \frac{1-\Psi_n}{3}\left(1+ \frac{\nu}{3}\sqrt{\frac{1-\Psi_n}{(\nu-1)(\nu-2)}} \right)\,, \qquad {\rm for} \quad |1-\Psi_n |\ll 1\,.
\end{align}

Detonation solutions can only exist in the narrow range $\alpha_{n}\in\left[\alpha_{n,\min}^{\rm det},\alpha_{n,\max}^{\rm det}\right]$. Models outside of this bound will not nucleate at all or give either deflagration or hybrid solutions (to be discussed below) or runaway walls, indicating that no static solution can be found. More realistically, in the latter case, we still expect the wall to reach a terminal velocity due to out-of-equilibrium effects which are not considered here. We refer the readers to Refs.~\cite{Bodeker:2009qy,Bodeker:2017cim,Hoche:2020ysm,Azatov:2020ufh, Gouttenoire:2021kjv} for discussions of the wall velocity in the ultrarelativistic limit.

\paragraph{Deflagrations}
For walls propagating at a subsonic speed ($\xi_w<c_b$), solutions correspond to deflagration profiles. These solutions are characterized by a shock wave propagating in front of the wall while the plasma is at rest behind it, which implies $v_-=\xi_w$. Furthermore, the fluid in front of the shock wave is unperturbed, leading to the conditions $v_{\rm sw,+}=\xi_{\rm sw}$ and $T_{\rm sw,+}=T_n$, where the subscript ``sw'' indicates that the quantity is defined in the neighborhood and the rest frame of the shock-wave front.   

The matching relations Eqs.~(\ref{eq:matching}) can be applied directly to the shock front, where they lead to simpler equations since the equation of state is the same behind and ahead of the shock wave. One can show that these matching conditions simplify to
\begin{subequations}
\label{eq:matchsw}
\begin{align}
    v_{\rm sw,-}&=\frac{c_{s}^2}{\xi_{\rm sw}}\,, \label{eq:matchsw1}\\
    \frac{\xi_{\rm sw}}{v_{\rm sw,-}} &= \frac{(\mu-1)T_{\rm sw,-}^\mu+T_n^\mu}{(\mu-1)T_n^\mu+T_{\rm sw,-}^\mu}\,.\label{eq:matchsw2}
\end{align}
\end{subequations}

Generally, determining the wall velocity of a deflagration is numerically more involved than for detonations as one needs to solve in addition the fluid equations~\eqref{eq:continuity2}; one has to integrate Eqs.~\eqref{eq:continuity2} through the shock wave (from $\xi_w$ to $\xi_{\rm sw}$) to relate the quantities right in front of the wall to those right behind the shock front. The boundary conditions for the fluid equations ($v_-=\xi_w$ and $T_{\rm sw,+}=T_n$) are given at two distinct locations so that the full initial conditions for these differential equations are not known \textit{a priori}. Therefore, it is typically necessary to use a shooting algorithm, varying $T_+$ and $\xi_w$ until all the matching equations~\eqref{eq:matchingEqs} and~\eqref{eq:matchsw} can be simultaneously satisfied. It should be noted here, that the obtained solution does not exactly satisfy the LTE condition Eq.~\eqref{eq:matchentropy} at the shock front, although numerically the deviations are small. For wall velocities of $\xi_w < 0.8$, we find that the deviation from Eq.~\eqref{eq:matchentropy} is smaller than $1\%$. Deviations become more sizeable for larger wall velocities, but we in any case expect out-of-equilibrium effects to become important in that regime~\cite{Laurent:2022jrs}.

Unlike for detonations, we observe that $\xi_w$ is an increasing function of $\alpha_{n}$, see Fig.~\ref{fig:vw}. Again, only a finite range of $\alpha_{n}$ can sustain a deflagration solution. The static wall limit ($\xi_w=0$) is attained when $\alpha_{n}=\alpha_{n,\min}^{\rm def}$. In this limit, we intuitively expect the fluid to be unperturbed by the presence of the wall, which implies $\xi_w=v_-=v_+=0$ and $T_-=T_+=T_n$. From Eq.~\eqref{eq:matching}, one easily sees that this situation corresponds to $Dp(T_n)=0$. Therefore, the condition $\xi_w>0$ (or $\alpha_{n}>\alpha_{n,\min}^{\rm def}$) is equivalent to $Dp(T_n)<0$, which is necessary for the bubble to be able to nucleate. From the definitions of $\alpha_{n}$ and $\Psi_n$ and using the condition $Dp(T_n)=0$, one can show that $\alpha_{n}$ cannot be smaller than $(1-\Psi_n)/3$. A similar condition is obtained by requiring the vacuum energy $\epsilon$ to be positive, which results in a lower bound of $\frac{1}{3\mu}(\mu-\nu)$. The minimal $\alpha_{n}$ is therefore given by
\begin{align}\label{eq:alphamin}
    \alpha_{n,\min}^{\rm def}=\max\left[ \frac{1-\Psi_n}{3},\frac{\mu-\nu}{3\mu}\right] \,.
\end{align}

\paragraph{Hybrids}

Finally, in the gap between deflagrations and detonations ($c_b\leq\xi_w<\xi_J$), phase transitions are described by hybrid solutions. These walls have both a rarefaction and a shock wave, and satisfy the boundary condition $v_-=c_b$. Since they also satisfy $v_-\geq v_+$, these solutions can technically be classified as deflagrations, and they indeed share similar properties with the latter. The wall velocity for hybrid solutions can be determined with a method completely analogous to deflagrations, with the appropriate boundary conditions.

Interestingly, the transition from deflagration to hybrid is continuous with respect to $\alpha_{n}$. Thus, in most applications, it is not necessary to make any distinction between the two. The highest wall velocity is reached at $\xi_w=\xi_J$ when $\alpha_{n}=\alpha_{n,\max}^{\rm hyb}$. This limit is obtained when the wall is infinitesimally close to the shock front. Any increase in $\alpha_{n}$ would therefore result in the wall getting ahead of the shock wave, which is not physically consistent. Since in this limit, the wall and shock front are essentially at the same location, it is not necessary to integrate Eqs.~\eqref{eq:continuity2} through the shock wave and the matching equations can be applied directly to relate the plasma behind the wall and in front of the shock wave. If entropy conservation were enforced at the shock front, this situation would be completely analogous to a Jouguet detonation. In reality, as discussed above, entropy is approximately conserved only when $T_{\rm sw,+} \approx T_{\rm sw,-}$, which is the case when $\Psi_n\approx 1$. Therefore, we conclude
\begin{align}
    \alpha_{n,\max}^{\rm hyb}\approx\alpha_{n,\max}^{\rm det}\,, \qquad {\rm for} \quad |1-\Psi_n |\ll 1\,.
\end{align}
We find numerically that for $\Psi_n$ not close to one, $\alpha_{n,\rm max}^{\rm hyb}$ is larger than $\alpha_{n,\rm max}^{\rm det}$.

\label{sec:model}

\subsection{Wall velocity}

An important consequence of the previous discussion is that regions of the parameter space with a detonation solution always come with a deflagration or hybrid solution as well. Specifically, since $\alpha_{n,\min}^{\rm def}$ is the smallest $\alpha_n$ at which a bubble can nucleate, and since $\alpha_{n,\max}^{\rm hyb}=\alpha_{n,\max}^{\rm det}$, the range of possible $\alpha_n$ for detonations is always included in the one for deflagrations and hybrids. If we imagine the bubbles to accelerate slowly to the steady-state solution after nucleation, we conclude that only the deflagration and hybrid solutions are relevant, as these are reached before the detonation solution would be obtained.

\begin{figure}[ht]
    \centering
    \includegraphics[width=0.6\linewidth]{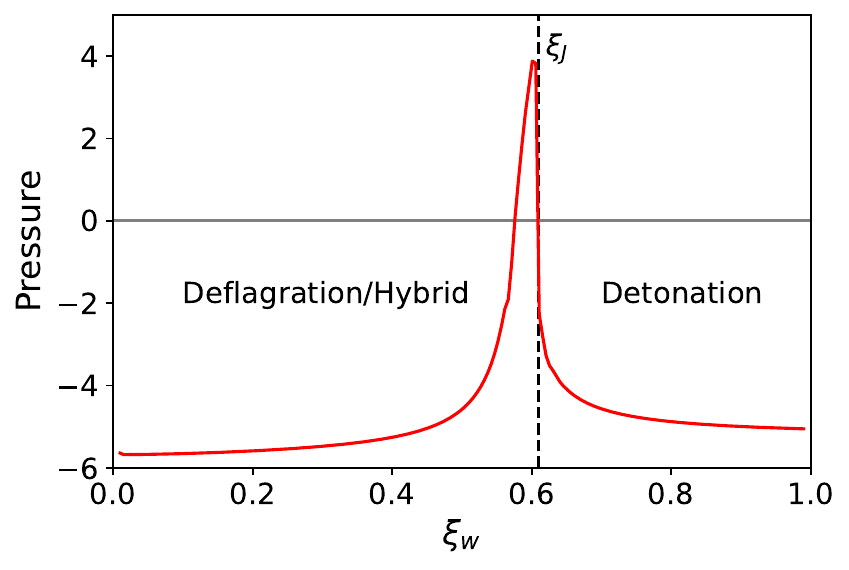}
    \caption{Example of the total pressure (in arbitrary units) acting on the wall, as defined in Eq.\ (39) of Ref.\ \cite{Laurent:2022jrs}.} 
    \label{fig:pressure}
\end{figure}

If, however, the early dynamics right after nucleation allows acceleration to supersonic speeds before stationary hydrodynamical behavior sets in, the wall could still reach the detonation solutions. It turns out that these detonation solutions are not stable. This is because the hydrodynamic LTE ``friction'' is a decreasing function of the wall velocity in the detonation regime~\cite{Ai:2021kak,Laurent:2022jrs}. For a detonation solution, any increase of the wall velocity would result in a smaller hydrodynamic ``friction'' which then cannot balance out the constant driving force due to the vacuum energy difference between the two phases, leading to unstable runaway behavior. In the deflagration and hybrid regime the force is an increasing function of the wall velocity and therefore the solutions are stable. This behavior is illustrated in Fig.\ \ref{fig:pressure}, where we see that the pressure is maximized at the Jouguet velocity. This pressure peak exists because of hydrodynamic effects which heat up the shock wave for hybrid solution. When the wall velocity exceeds the Jouguet velocity, the shock wave disappears and these hydrodynamic effects cease to exist, thereby decreasing the total pressure on the wall. We thus conclude that detonation solutions in LTE are not physical and we therefore omit them in this section.

\begin{figure}[ht]
    \centering
    \includegraphics[width=0.49\linewidth]{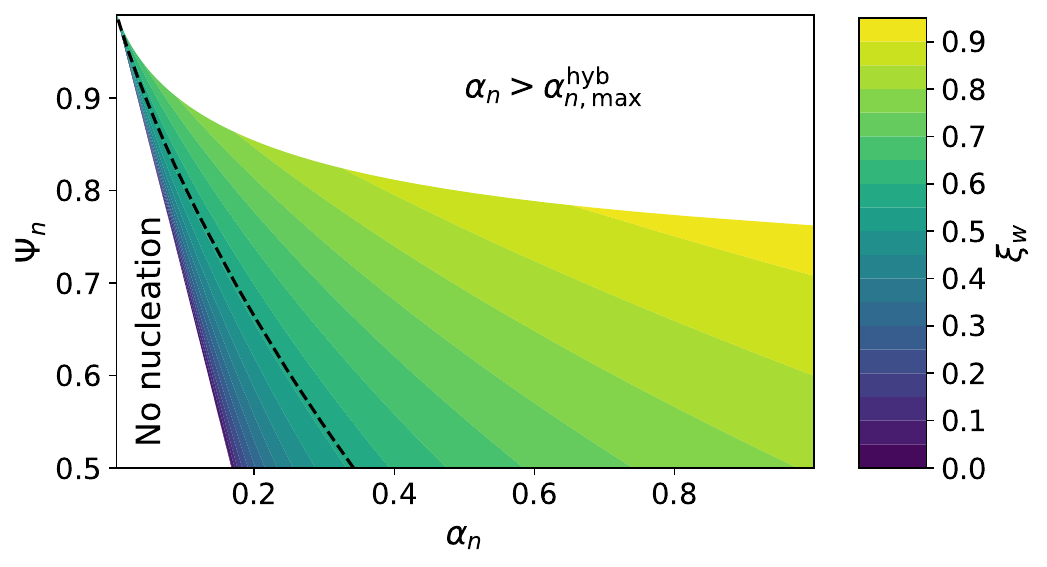}\hspace{0.01\linewidth}\includegraphics[width=0.49\linewidth]{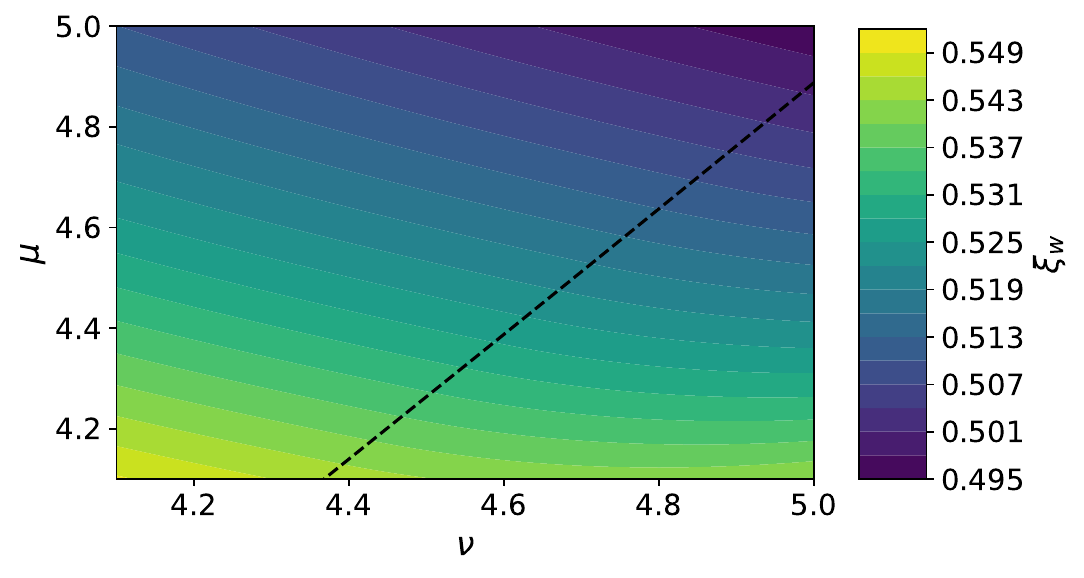}\\
    (a)\hspace{0.5\linewidth}(b)
    \caption{Wall velocity for deflagration and hybrid solutions as a function of: (a) $\alpha_n$ and $\Psi_n$ with $\nu=4.2$ and $\mu=4.1$; (b) $\nu$ and $\mu$ with $\alpha_n=0.1$ and $\Psi_n=0.8$. The black dashed line corresponds to $\xi_w=c_b$, such that deflagration and hybrid solutions are on the left and right sides of the line, respectively.}
    \label{fig:vw}
\end{figure}

Plots of the wall velocity for deflagration and hybrid solutions are shown in Fig.~\ref{fig:vw}. On the left-hand side, one can see that even for hybrid solutions, the wall can nearly reach the speed of light, provided that $\Psi_n$ is not too large and $\alpha_n$ is close to $\alpha_{n,\max}^{\rm hyb}$. This happens because, at low $\Psi_n$, $\xi_w$ and $\xi_J$ increase at a similar rate with $\alpha_n$. This has the interesting consequence that for $\Psi_n\lsim 0.8$, it is almost guaranteed to obtain a deflagration or hybrid solution, as $\alpha_{n,\max}^{\rm hyb}$ becomes very large which means most realistic models will satisfy $\alpha_n<\alpha_{n,\max}^{\rm hyb}$.\footnote{We remind the reader that GW simulations from sound waves have been performed only for $\alpha_n \lesssim 0.1$~\cite{Hindmarsh:2013xza,Hindmarsh:2015qta,Hindmarsh:2017gnf,Jinno:2020eqg,Jinno:2022mie}, except for Ref.~\cite{Cutting:2019zws}, where values up to $\alpha_n \lesssim 1$ were simulated. For those relatively large values of $\alpha_n$ the GW spectrum gets suppressed compared to the case of weaker phase transitions, especially for deflagrations.} On the other hand, for $\Psi_n$ closer to 1, the range of allowed $\alpha_n$ becomes narrower. This is consistent with the results of Ref.~\cite{Laurent:2022jrs}, who studied the singlet scalar extension for which $\Psi_n\gtrsim 0.9$. Another striking result of Ref.~\cite{Laurent:2022jrs} was the absence of walls with low velocity. Here, we do get velocities as low as $\xi_w=0$. But from the discussion above, we now understand that these slow walls correspond to $D p\approx 0$, where bubble nucleation is highly inefficient. Therefore, in a realistic situation where we assume that bubbles have been able to nucleate with a sufficiently large rate, $D p$ (and $\alpha_n$) cannot be arbitrarily small which leads to relatively large $\xi_w$, implying that the limit of $\xi_w \rightarrow 0$ in Fig. \ref{fig:vw} may not be realistic. 

The dependence of the wall velocity on the speeds of sound is shown in Fig.~\ref{fig:vw}b. The wall velocity depends more strongly on $c_s$, the speed of sound in front of the wall, and only weakly on $c_b$. Overall, we observe variations of order 20\% when varying $\mu$ and $\nu$ between 4 and 5 which can have a significant impact when high accuracy is needed.

Until now, we have not discussed the possibility of phase transitions with $\alpha_n>\alpha_{n,\max}^{\rm hyb}$. For these models, the LTE approximation predicts that no static deflagration, hybrid or detonation solution exists. Thus, the only possibility in this context is a nonstatic solution, which corresponds to a runaway wall. These walls never reach a terminal velocity and continue accelerating towards $\xi_w\to 1$ until the phase transition is completed. In reality however, runaway walls appear unlikely, as Ref.~\cite{Bodeker:2017cim} showed that the pressure created by interactions between the wall and gauge bosons diverges in the limit $\xi_w\to 1$ (we refer the readers to Refs.~\cite{Hoche:2020ysm,Azatov:2020ufh,Gouttenoire:2021kjv} for more recent discussions of this effect). This signals the breakdown of the LTE approximation at ultrarelativistic speeds, which cannot model such interactions as they create large deviations from equilibrium. Phase transitions can therefore be classified into two groups: deflagrations/hybrids with $\alpha_n<\alpha_{n,\max}^{\rm hyb}$ which can be treated approximately with the LTE approximation, and ultrarelativistic detonations with\footnote{We expect that the exact value of $\alpha_{n,\max}^{\rm hyb}$ will change once out-of-equilibrium effects are taken into account.} $\alpha_n>\alpha_{n,\max}^{\rm hyb}$ which are stopped at high velocity by out-of-equilibrium effects and thus cannot be modelled by LTE~\cite{Laurent:2022jrs}. 

\paragraph{Validity of the LTE approximation.}

One of the main sources of uncertainty in the calculation made in this paper stems from the assumption of LTE. Of course, in a realistic situation, this approximation is not exactly satisfied as the wall perturbs the plasma and can provoke a deviation from equilibrium around it. Typically, these perturbations create an additional source of friction slowing down the wall, making the actual wall velocity smaller than what is predicted by the LTE assumption.

Nevertheless, one can hope that these deviations from equilibrium are small and that LTE accurately represents the plasma. This is effectively the conclusion of Ref.~\cite{Laurent:2022jrs}, who studied the wall velocity in the singlet scalar extension and found a good agreement between LTE and the full treatment including deviations from equilibrium of the top quark. More recently, Ref.~\cite{DeCurtis:2023hil} performed a similar analysis where the authors observed a greater sensitivity on the out-of-equilibrium perturbations. Still, the wall velocities found with the two treatments only deviate by approximately 20\%.

The model used in these two studies only explores the region of parameter space where $\Psi_n\gsim 0.9$. Therefore, the parameter space $\Psi_n<0.9$ has not yet been tested, and it remains to be seen if LTE can accurately describe the plasma in this region. In general, we expect the out-of-equilibrium effects to become more important at higher $\alpha_n$ and $\xi_w$. Furthermore, the magnitude of the out-of-equilibrium effects can also depend on model-dependent features, like the variation of the vacuum expectation values of the scalar fields. Thus, it is impossible to guarantee the applicability of LTE with only the model-independent description made in this paper.

Finally, even if the deviation from equilibrium turns out to be large, LTE might still be a useful tool as it offers an upper bound for the wall velocity. An interesting consequence is that if a deflagration or hybrid solution exists within LTE, it will also exist when the out-of-equilibrium effects are considered. In other words, LTE always underestimates $\alpha_{n,\min}^{\rm hyb}$. In particular, models with $\Psi_n\lsim 0.75$ cannot become a detonation, as LTE predicts a prohibitively large (possibly infinite) $\alpha_{n,\max}^{\rm hyb}$ in this region. 

\paragraph{Fit of the wall velocity for $\nu=\mu=4$.}

Most phase transitions of interest happen when the Universe is radiation dominated. In that case, it is generally a good approximation to assume the speeds of sound to be $c_{b,s}^2\approx 1/3$. For the reader's convenience, we present here a numerical fit for $\xi_w$ that is valid in this restricted region of the parameter space.

It is possible to derive analytically a formula for the wall velocity when $\alpha_n\approx \alpha_{n,\min}^{\rm def}$, which corresponds to small $\xi_w$. It can be obtained by expanding linearly the fluid equations \eqref{eq:continuity2} and \eqref{eq:matchingEqs} for small $\xi_w$, $v_+$ and $v_-$. The solution then reads
\begin{align}
    \xi_w^{\rm low}= \sqrt{\frac{3\alpha_n+\Psi_n-1}{2(2-3\Psi_n+\Psi^3_n)}}\,.
\end{align}
This formula is accurate for $\xi_w\lsim 0.5$.

Similarly, one can expand the fluid equations for $\xi_w\to 1$ to get a formula valid at high velocity. It can be shown that $1-\xi_w$ is then proportional to $1/\alpha_n$. This motivates the following formula:
\begin{align}
    \xi_w^{\rm high} = \xi_J\left( 1 - a\frac{(1-\Psi_n)^b}{\alpha_n}\right)\,,
\end{align}
where $\xi_J$ is the Jouguet velocity defined in Eq.~\eqref{eq:vJ}, and $a$ and $b$ are numerical coefficients that need to be fitted against some data.

One can finally obtain a fit valid for all wall velocities by interpolating between $\xi_w^{\rm low}$ and $\xi_w^{\rm high}$. We achieve this by taking the $p$-norm of these two formulas, which is defined by 
\begin{align}\label{eq:fit}
    \xi_w^{\rm fit} = \| (\xi_w^{\rm low},\xi_w^{\rm high})\|_p \equiv \left( \left| \xi_w^{\rm low}\right| ^p+\left| \xi_w^{\rm high}\right|^p\right)^{1/p}\,,
\end{align}
with $p<0$. In the limit $p\to -\infty$, the $p$-norm converges towards the minimum of the two arguments. So for finite negative $p$, it can be interpreted as a smoothed minimum function that can be used to interpolate between the low and high $\xi_w$ regime efficiently.

\begin{figure}[t]
    \centering
    \includegraphics[width=0.6\linewidth]{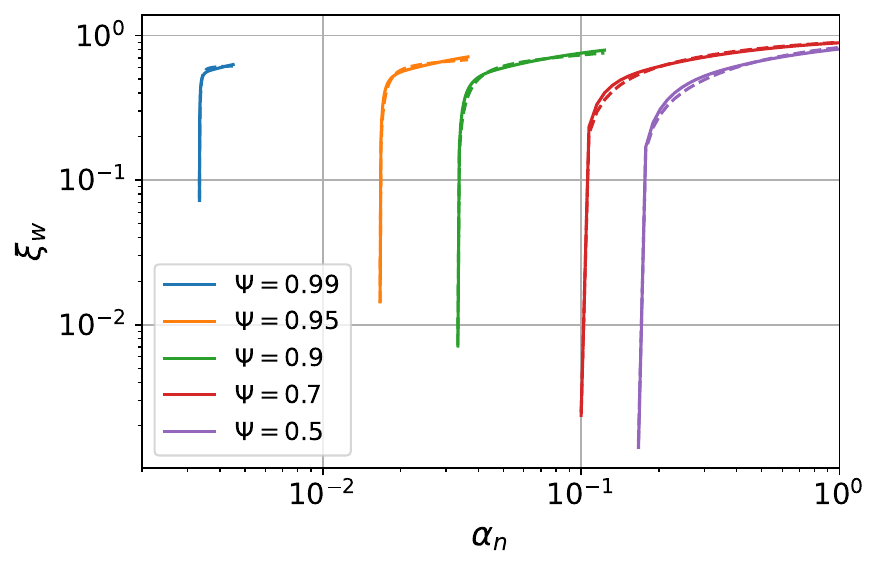}
    \caption{Comparison of the numerical fit \eqref{eq:fit} (dashed lines) to the complete numerical solution (solid lines) for several values of $\Psi$.}
    \label{fig:fit}
\end{figure}

We have fitted the formula \eqref{eq:fit} against the data shown in Fig.~\ref{fig:vw} and found the optimal numerical coefficients to be $a=0.2233$, $b=1.704$ and $p=-3.433$. These values give an error well below 10\% most of the times. Note that the fit should only be applied for $\alpha_n>\alpha^{\rm def}_{n,\rm min}$ and the resulting wall velocity should be smaller than the Jouguet velocity, e.g., $\xi^{\rm fit}_w < \xi_J$.
A comparison of the fit to the complete numerical solution is shown in Fig.~\ref{fig:fit}. 

\subsection{Kinetic energy fraction}

Now that we have a way to compute the wall velocity, it becomes possible to determine the kinetic energy fraction $K$ completely from the four parameters $\alpha_n$, $\Psi_n$, $c_s$ and $c_b$. This is a great improvement over previous studies that had to compute $K$ as a function of $\xi_w$, which had to remain unspecified.

We introduced the adiabatic index in Sec.~\ref{sec:hydro} as
\begin{equation}
    \Gamma = \frac{\omega_n}{e_n}\,.
\end{equation}
In the context of the model-independent parameterization of the hydrodynamics in terms of $\alpha_n,c_b,c_s$ and $\Psi_n$, the adiabatic index may be understood as a fifth parameter that is needed to capture the details of the model. However, we demonstrate here how it can be obtained from the other four parameters. This is only possible if we explicitly rely on the assumption that the speed of sound is approximately constant in the broken phase, i.e.,
\begin{equation}
    c_b^2 \approx \frac{p_b(T_n)}{e_b(T_n)}\,.
\end{equation}
Now $\alpha_n$ simplifies to
\begin{equation}
\label{eq:B54}
    \alpha_n \approx \frac{e_n - \frac{1}{c_b^2} p_n}{3\omega_n} = \frac{1}{3\Gamma}\left(1+ \frac{1}{c_b^2}\right)- \frac{1}{3 c_b^2}\,,
\end{equation}
and we thus approximate the adiabatic index as
\begin{equation}
    \Gamma \approx \frac{\nu}{3 \alpha_n + \nu -1}\,.
\end{equation}
The above relation becomes exact in the template model.

\begin{figure}[ht]
    \centering
    \includegraphics[width=0.49\linewidth]{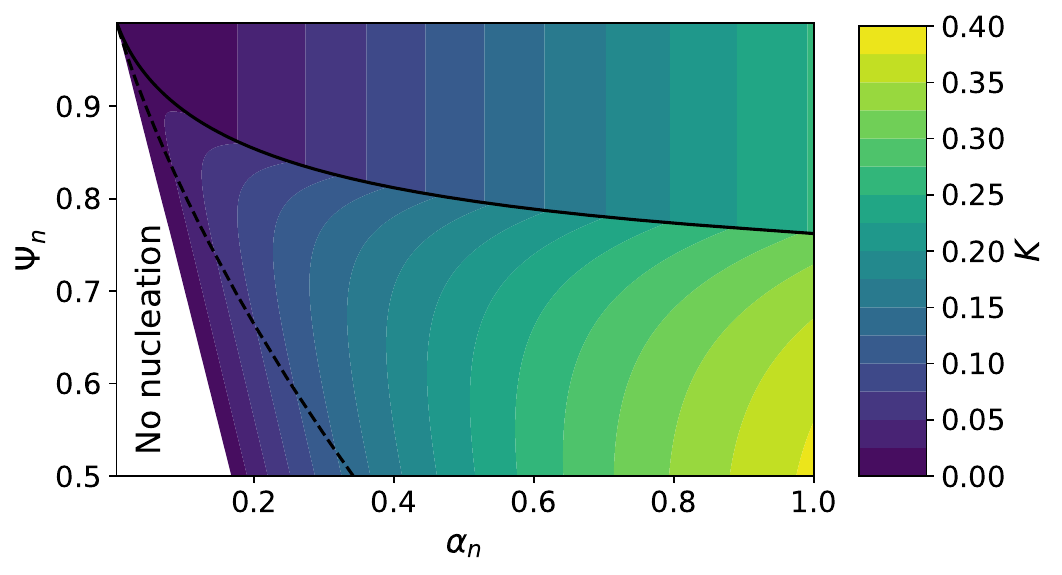}\hspace{0.01\linewidth}\includegraphics[width=0.49\linewidth]{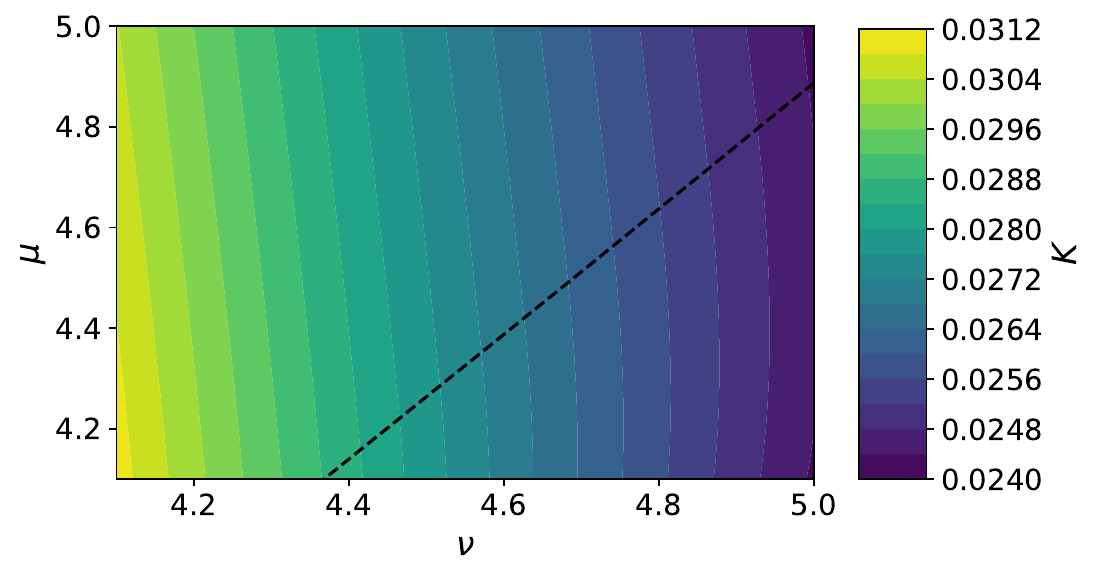}\\[-2mm]
    (a)\hspace{0.48\linewidth}(b) \\[2mm]
    \includegraphics[width=0.5\linewidth]{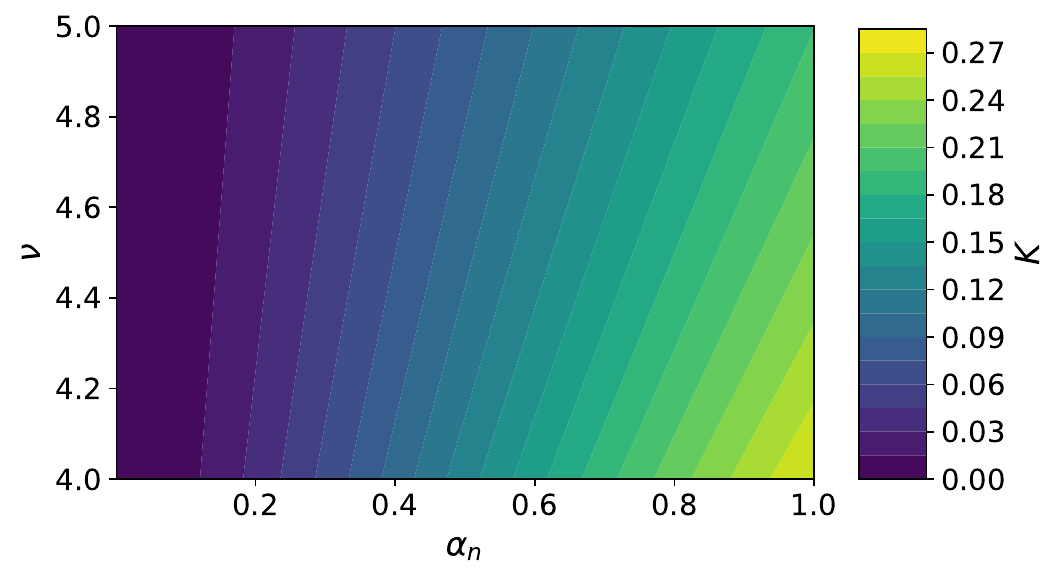}\\[-2mm]
    (c)
    \caption{(a) and (b): Kinetic energy fraction for the same parameters as used in Fig.~\ref{fig:vw}. The black dashed line corresponds to $\xi_w=c_b$, while the black solid line separates hybrid walls from ultrarelativistic detonations. (c): Kinetic energy fraction for ultrarelativistic detonations only as a function of $\alpha_n$ and $\nu$ (we do not specify $\Psi_n$ or $\mu$ as they have no effect on $K$ in that regime).}
    \label{fig:K}
\end{figure}

Plots of the kinetic energy fraction are shown in Fig.~\ref{fig:K}. Unlike the calculation of the wall velocity, it is still possible to estimate $K$ even when $\alpha_n>\alpha_{n,\max}^{\rm hyb}$. Effectively, as argued in the previous subsection, these walls become ultrarelativistic detonation solutions satisfying $\gamma_w\gg 1$. It was pointed out in Ref.~\cite{Espinosa:2010hh} that the kinetic energy fraction does not depend on $\xi_w$ in this limit; it can therefore be computed even if the precise value of $\xi_w$ is unknown, and therefore we do not need to rely on LTE approximation. 

As expected, the transition from hybrid to detonation solutions is not continuous in $K$, since the wall velocity suddenly jumps from $\xi_J$ to 1. We generally observe that $K$ is greater for hybrid walls than for detonations. Another interesting result is that, for hybrid walls, $K$ is not a strictly increasing function of $\alpha_n$. Effectively, when $\Psi_n\gtrsim 0.9$, most of the kinetic energy is stored in the shock wave, which becomes thinner as $\alpha_n$ approaches $\alpha_{n,\max}^{\rm hyb}$, leading to a smaller $K$. For smaller $\Psi_n$, the energy in the rarefaction wave also becomes important and thus can compensate for the thinner shock wave, and therefore $K$ increases with $\alpha_n$, as expected. The kinetic energy fraction for detonations is always an increasing function of $\alpha_n$ and $c_b$.

The dependence of $K$ on the sound velocities for deflagration and hybrid walls is shown in Fig.~\ref{fig:K}b. Contrary to $\xi_w$, $K$ depends mostly on $c_b$ and very little on $c_s$. Again, variations of $\mu$ and $\nu$ between 4 and 5 only have a quantitative impact on $K$, causing a relative change of order 25\%.

\section{Conclusions}
\label{sec:Conc}

FOPTs are a promising source of GW signals that could be observable with the next generation of GW detectors~\cite{Caprini:2015zlo,Caprini:2018mtu,Caprini:2019egz,LISACosmologyWorkingGroup:2022jok}. These transitions are usually associated with new physics beyond the Standard Model, and detecting the associated GW signal would be a triumph of particle physics. Therefore, the properties of the generated GW signals have been extensively studied in recent years. The GW power spectrum is sensitive to six quantities, namely the Hubble parameter during percolation $H_*$, the phase transition strength $\alpha_n$, the bubble wall velocity $\xi_w$, the transition rate $\beta$, and the sound speeds $c_s$ and $c_b$.\footnote{In the simplest bag equation of state, the sound speeds are fixed at $1/\sqrt{3}$ and are therefore not treated as parameters.} Among these quantities, the wall velocity is the most difficult to compute and is usually model dependent. Often, the value of the wall velocity is either treated as an unknown parameter or approximated as $\xi_w \rightarrow 1$.

In this work, we have shown that the uncertainty associated to the wall velocity can be eliminated if the plasma is assumed to be in LTE, which has been shown to be a good approximation in the singlet scalar extension of the Standard Model in Ref.~\cite{Laurent:2020gpg}. The benefit of the LTE approximation is that entropy conservation provides a new matching condition which makes a determination of the wall velocity possible, as was shown in Ref.~\cite{Ai:2021kak} in the bag equation of state.   In this paper, we have demonstrated that this new matching condition can be used to compute the bubble wall velocity and the energy fraction in more general situations.  We found that the dependence on the wall velocity would be replaced by a dependence on a new thermodynamical quantity $\Psi_n$, the ratio of the enthalpies in the broken and symmetric phases. We have demonstrated that the hydrodynamics in LTE can be fully characterized by four quantities: $\Psi_n$, $\alpha_n$, $c_s$, and $c_b$. 

We have considered all the possible modes of fluid motion: detonations, hybrids, and deflagrations, without resorting to the planar wall approximation. 
We found that, whenever the parameters allow for a detonation solution, a deflagration or hybrid solution also exists, rendering the detonation solutions irrelevant in LTE. This indicates that the determination of the wall velocity for detonations requires the inclusion of out-of-equilibrium effects. For the deflagration and hybrid solutions, our results for the bubble wall velocities should provide a good approximation, and at least an upper bound, of the velocities when calculated with out-of-equilibrium effects included. 

We found that the wall velocity is an increasing function of both $\alpha_n$ and $\Psi_n$. When $\alpha_n$ is varied with the other parameters fixed, a solution exists only within a finite range of $\alpha_n$. Outside of this range,  bubbles do not nucleate, or significant out-of-equilibrium effects are expected to violate LTE. The allowed range of $\alpha_n$ is narrow for $\Psi_n$ close to unity, but becomes broader when $\Psi_n$ decreases. For $\Psi_n\lesssim 0.8$, the solution exists over a large range of $\alpha_n$, suggesting that the LTE approximation can potentially have a broad range of applications, although finding a solution itself does not indicate that LTE is applicable.    Notably, we observed that the wall velocity can approach the speed of light when $\Psi_n$ is not too large. The sound speeds also affect the wall velocity, with the sound speed in the symmetric phase $c_s$ having a greater impact than that in the broken phase $c_b$. We found that varying $c_s$ and $c_b$ between $1/\sqrt{3}$ and $1/\sqrt{4}$ resulted in variations of around 20\% in the wall velocity. Since there are many models in which the phase transition would occur with $c_{s,b} \approx 1/\sqrt{3}$, we have provided a fit for the wall velocity in the special case $c_{s,b}=1/\sqrt{3}$.

We have also computed the kinetic energy fraction $K$ for both our LTE solutions and ultrarelativistic walls. We confirmed that hybrid walls generally have a greater $K$ than detonations. We also found an interesting non-monotonic behavior of $K$ as a function of $\alpha_n$, resulting from a competition between the kinetic energy stored in the shock wave and the rarefaction wave. The energy fraction $K$ is also sensitive to the sound speeds, but mostly to $c_b$ and only slightly to $c_s$. We have observed variations of approximately $25\%$ when varying $c_s$ and $c_b$ between $1/\sqrt{3}$ and $1/\sqrt{4}$. 

At last, we have provided a Python snippet which the reader can use to compute the wall velocity and energy fraction for other values of interest of the parameters $(\alpha_n,\Psi_n,c_s,c_b)$. Even though for some regions of the parameters, unexpected out-of-equilibrium effects may have significant contributions to the friction, our results on the wall velocity can still be interpreted as an upper bound as out-of-equilibrium effects always make the wall slower compared with the case when they are absent.

\section*{Acknowledgments}
We would like to express our gratitude to Kavli IPMU, where this collaboration was initiated at the workshop ``What the heck happens when the Universe boils?''. We would also like to thank all the participants for the helpful discussions during that workshop. WYA is supported by the UK Engineering and Physical Sciences Research Council (EPSRC), under Research Grant No. EP/V002821/1. BL is supported by the Fonds de recherche du Québec Nature et technologies (FRQNT). JvdV is supported by the Dutch Research Council (NWO), under project  number VI.Veni.212.133.

\newpage
\begin{appendix}
\renewcommand{\theequation}{\Alph{section}\arabic{equation}}

\section{Entropy conservation in local equilibrium}\label{ap:LTEEntropy}

In this Appendix, we show that entropy is conserved in LTE. From the argument given below, one can see that in principle the bubble wall velocity calculated with the equation of motion for $\phi$, in the case of LTE, can be recovered from a calculation with the new matching condition studied in Ref.~\cite{Ai:2021kak} and in the present work.

The energy-momentum tensors for the scalar background and the fluid read
\begin{align}
T_{\phi}^{\mu\nu}&=(\partial^\mu\phi)\partial^\nu\phi-g^{\mu\nu}\left(\frac{1}{2}(\partial\phi)^2-V(\phi)\right),\\
T_f^{\mu\nu}&=(e_f+p_f)u^\mu u^\nu-p_f g^{\mu\nu}\,,
\end{align}
where $V(\phi)$ is the zero-temperature tree-level potential for the scalar field and $u^\mu$ is the normalized four-velocity field ($u^\mu u_\mu=1$) for the fluid. Here, $e_f$ and $p_f$ are the fluid energy density and pressure which vanish as the temperature approaches zero. However, one usually combines the fluid energy density and pressure with the tree-level scalar potential energy, $e=e_f+V(\phi)$, $p=p_f-V(\phi)$. $e$ and $p$ do not vanish at zero temperature, but the advantage of using them is that in terms of $e$ and $p$ the matching conditions for hydrodynamic quantities take the form of Eqs.~\eqref{eq:matching} that appear commonly in the literature. Note that the fluid enthalpy $\omega=e_f+p_f=e+p$. In terms of $e$ and $p$, the energy-momentum tensor for the fluid then takes the form as given in Eq.~\eqref{eq:emfluid}.

The partial derivatives of the two separate components of the energy-momentum tensors are given by
\begin{align}
\partial_\mu T_\phi^{\mu\nu}=
(\partial^\nu\phi)\left[\partial^2\phi+V'(\phi)\right]\,,
\end{align}
and 
\begin{align}
\partial_\mu T^{\mu\nu}_f=
\partial_\mu[\omega u^\mu u^\nu]-(\partial^\nu T)\frac{\partial p}{\partial T}+ (\partial^\nu\phi) V'_1(\phi,T)\,,
\end{align}
where $V_1(\phi,T)=-p_f=-(p+V(\phi))$. Energy-momentum conservation of the full energy-momentum tensor then gives 
\begin{align}
(\partial^\nu\phi)[\partial^2\phi+V'_{\rm eff}(\phi,T)]+\partial_\mu[\omega u^\mu u^\nu]-(\partial^\nu T)\frac{\partial p}{\partial T}=0\,,
\end{align}
where $V_{\rm eff}(\phi,T)=V(\phi)+V_1(\phi,T)$. Multiplying the above equation with $u_\nu$, we obtain
\begin{align}
0&=(u_\nu \partial^\nu\phi)\left[\partial^2\phi+V'_{\rm eff}(\phi,T)\right]+u_\nu\partial_\mu(Ts u^\mu u^\nu)-u_\nu(\partial^\nu T)\frac{\partial p}{\partial T}\notag\\
&=(u_\nu \partial^\nu\phi)\left[\partial^2\phi+V'_{\rm eff}(\phi,T)\right]+T\partial\cdot S\,,\label{eq:consEM}
\end{align}
where we defined the entropy current $S^\mu= s u^\mu$. We also used $u^\mu u_\mu=1$ and (thus) $u_\mu \partial_\nu u^\mu=0$.

The general equation of motion for the scalar background is 
\begin{align}
\partial^2\phi+V_{\rm eff}'(\phi)+\sum_i\frac{d m^2_i(\phi)}{d\phi}\int\frac{d^3\vec{p}}{(2\pi)^3 2E_i}\delta f_i(p,x)=0\,.
\end{align}
LTE can then be defined with the condition
\begin{align}
\delta f_i=0\ {\rm for\ all\ } i\,,
\end{align}
such that the scalar equation of motion becomes 
\begin{align}
\partial^2\phi+V_{\rm eff}'(\phi)=0\,.
\end{align}
Plugging this equation of motion into Eq.~\eqref{eq:consEM}, we see that LTE is equivalent to
\begin{align}
\label{eq:entropy_conser}
\partial\cdot S=0\,.
\end{align}

\section{Uniform speed of sound approximation}\label{ap:sound}

In this section, we derive the matching equations \eqref{eq:matchingEqs} in two regimes where the speed of sound is approximately uniform in the plasma. The purpose of this approximation is to write the matching equations in a model-independent way and to eliminate the dependence on $T_-$. This allows for the use of our results when only the nucleation temperature is known.

For the time being, however, we will keep the discussion as general as possible and avoid making the assumption of uniform sound speed. A quantity that will turn out to be of particular importance is
\begin{align}
\label{eq:v_app}
    \frac{\delta p}{\delta e}=\frac{p_b(T_+)-p_b(T_-)}{e_b(T_+)-e_b(T_-)}\equiv \zeta^2(T_+,T_-)\,, 
\end{align}
which is, at this point, a function of $T_-$ and $T_+$.
After some algebra, one can obtain from Eqs.~\eqref{eq:matchingdelta}
\begin{subequations}
\begin{align}
\label{eq:match0_app}
    &\frac{v_+}{v_-}=\frac{\left(\frac{v_+v_-}{\zeta^2}-1\right)+3\alpha_+  }{\left(\frac{v_+v_-}{\zeta^2}-1\right)+3v_+v_-\alpha_+}\,, \\
\label{eq:match1_app}
    &\frac{\Delta p}{3\omega_+}\left(1-\frac{v_+ v_-}{\zeta^2}\right)=v_+v_-\alpha_{+}\,,
\end{align}
\end{subequations}
with $\alpha_+$ defined as in Eq.~\eqref{eq:defalpha}. These equations depend explicitly on $T_-$ and $T_+$,  which makes them unsuitable for a model-independent application. Below, we will discuss regimes in which one can actually eliminate the $T_-$-dependence in $\zeta$. After doing that, $T_-$ would be completely eliminated in Eq.~\eqref{eq:match0_app}.  

Let us now extend the analysis above with the new matching condition Eq.~\eqref{eq:matchentropy}. 
Substituting it into Eq.~\eqref{eq:match1_app}, one obtains
\begin{align}
\label{eq:match2_app}
    \frac{p_s(T_+)-p_b(T_+\gamma_+/\gamma_-)}{3\omega_+}\left(1-\frac{v_+ v_-}{\zeta^2}\right)=v_+v_-\alpha_{+}\,.
\end{align}
The above equation is still not so useful for a model-independent analysis because it involves explicit expressions of the pressure, which equivalently require one to specify the effective potential. In analogy with Refs.~\cite{Giese:2020rtr,Giese:2020znk}, we will now eliminate the explicit dependence on the pressure, and express everything in terms of a few thermodynamics quantities evaluated at a single temperature. Surprisingly, in the regimes where we can eliminate the $T_-$-dependence in $\zeta$, we can also eliminate the explicit dependence on the pressure in Eq.~\eqref{eq:match2_app} as we now show.

\paragraph{Regime $(i)$: Approximately temperature-independent sound speed in the broken phase}

If the sound speed in the broken phase is approximately temperature independent, then the equation of state must follow the template model, in which we have 
\begin{align}
    \frac{\delta p_b(T)}{\delta e_b(T)}\approx \frac{1}{\nu-1} = c^2_b\equiv c^2_b(T_n)\,,
\end{align}
Substituting the above relation into Eq.~\eqref{eq:v_app}, one obtains 
\begin{align}
    \zeta(T_+,T_-)\approx c_b\,.
\end{align}
and substituting into Eq.~\eqref{eq:match0_app} gives
\begin{align}
    \label{eq:matchingvpvm_app}
    \frac{v_+}{v_-}\approx \frac{v_+v_-(\nu-1)-1+3\alpha_+  }{v_+v_-(\nu-1)-1+3v_+v_-\alpha_+}\,,
\end{align}
which is equivalent to Eq.~\eqref{eq:matchingEqsA}.

With the assumption that the sound speed in the broken phase is approximately constant, one should have~\cite{Leitao:2014pda} 
\begin{align}
    e_b(T)\approx \frac{1}{3}a_- (\nu-1) T^\nu\,,\quad p_b(T)\approx \frac{1}{3}a_- T^\nu \,.
\end{align}
Using the definitions of $\alpha_+$ and $\Psi_+$, one can show that
\begin{subequations}
\begin{align}
   &e_s(T_+)\approx (3\alpha_+ +\nu-1)\frac{a_-T_+^\nu}{3\Psi_+}\,,\\
&p_s(T_+)\approx (1-3\alpha_+)\frac{a_-T_+^\nu}{3\Psi_+}\,. 
\end{align}
\end{subequations}
Substituting the above equations into Eq.~\eqref{eq:match2_app}, one obtains
\begin{align}
\label{eq:match22_app}
    \frac{-\left(\frac{\gamma_+}{\gamma_-}\right)^{\nu} \Psi_+ + 1 - 3 \alpha_{+} }{3 \nu } \left[1-(\nu-1) v_+ v_-\right]\approx v_+v_-\alpha_{+}\,,
\end{align}
which is Eq.~\eqref{eq:matchingEqsB}. Note that the above derivation does not assume the equation of state in the symmetric phase.

\paragraph{Regime $(ii)$: Linear regime for the temperatures} 
Another regime is when $T_+\approx T_-$. In this case, one can then linearize $X_b(T_-)$ around $T_+$ (where $X$ denotes any thermodynamic quantity), and $\delta X$ can be approximated by 
\begin{align}
   \delta X\approx (T_+-T_-)\left.\frac{d X_b(T)}{d T}\right|_{T_+}\,, 
\end{align}
and, correspondingly, the ratio $\delta p/\delta e$ can be approximated as
\begin{equation}
\label{eq:sosdef_app}
    \zeta(T_+,T_-) \approx \frac{dp_b/dT}{d e_b/dT}\bigg\vert_{T_+} \equiv c_{b,+}^2 \equiv\frac{1}{\nu_+-1}\,,
\end{equation}
where $c_{b,+}$ denotes the sound speed in the broken phase evaluated at $T_+$. Substituting the above into Eq.~\eqref{eq:match0_app}, one obtains the same equation as~\eqref{eq:matchingvpvm_app} except that $\nu$ is replaced with $\nu_+$. 

To eliminate the dependence on the pressure in Eq.~\eqref{eq:match2_app}, we expand the pressure $p_b$ around its value at $T_+$. Since we eliminated $T_-$ using the matching condition Eq.~\eqref{eq:matchentropy}, the expansion around $T_- \approx T_+$ is now an expansion around $\gamma_+/\gamma_- \approx 1$. Expanding $p_b(T_+\gamma_+/\gamma_-)$ up to second order, we have 
\begin{align}
\label{eq:newmatchingapprox_app}
    &\frac{p_s(T_+)-p_b\left(\frac{T_+\gamma_+}{\gamma_-}\right)}{3\omega_+}\  \nonumber \\ \ &  \qquad = \frac{1}{3\omega_+}\Bigg(p_s(T_+) - p_b(T_+) -\omega_b(T_+)\left(\frac{\gamma_+}{\gamma_-}-1\right) 
    -\frac{1}{2} \omega_b(T_+)(\nu_+ -1)\left(\frac{\gamma_+}{\gamma_-}-1\right)^2 \Bigg)  +\cdots\nonumber\\
    \ & \qquad = \frac{1}{3\omega_+}\left( \frac{3\omega_+ \alpha_{+} -D e}{1- \nu_+ } - \omega_b(T_+)\frac{1}{ \nu_+ }\left(\frac{\gamma_+}{\gamma_-}\right)^{ \nu_+ } + \frac{\omega_b(T_+)}{ \nu_+ }  \right)+\cdots \nonumber \\
    \ & \qquad =  \frac{-\left(\frac{\gamma_+}{\gamma_-}\right)^{ \nu_+ } \Psi_++ 1 - 3 \alpha_{+} }{3  \nu_+ } +\cdots\,,
\end{align}
where the dots denote terms of order $\mathcal O (\gamma_+/\gamma_-)^3$ and we have used the relations
\begin{align}
    \omega=e+p=T\frac{d p}{d T}\quad\Rightarrow\quad \frac{d e}{d T}=T\frac{d^2 p}{d T^2} \,,
\end{align}
and the definition of the sound speed. 
With Eq.~\eqref{eq:newmatchingapprox_app}, one gets the same equation as~\eqref{eq:match22_app} except that $\nu$ is replaced by $\nu_+$. Furthermore, if one adds the condition $T_+\approx T_n$ which is generally the case when $T_+\approx T_-$, $\nu_+$ can safely be replaced by $\nu$, recovering the same equations as in the first regime. Note that in the derivation of the matching equations in this regime one does not assume the equations of state in either phase.

In both regimes, we see that once we know the parameters $\alpha_+$, $\Psi_+$ and $\nu_+$ (or $\nu$), one can solve $v_+$ and $v_-$ from Eqs.~\eqref{eq:matchingvpvm_app} and~\eqref{eq:match22_app}, determining the wall velocity $\xi_w$.

\section{Code snippet}\label{ap:code}

In this section, we briefly describe the algorithm used to solve the hydrodynamic equations and provide a snippet of Python code that computes $\xi_w$ and the efficiency factor $\kappa$ in the template model. We have tested the code snippet with Python version 3.9. We have tested the code for values of $0.5<\Psi_n<0.99$ and $1/10 < c_{s,b}^2 < 2/3$. In some of these cases the code generates a warning due to the high level of precision that is requested. We have however confirmed that the value of the wall velocity is correct, so this warning can safely be ignored.

We start by describing how we compute the wall velocity for deflagration and hybrid walls, for which $v_-=\min(\xi_w,c_b)$. As the temperature in front of the wall is unknown at first, one needs to integrate Eqs.~\eqref{eq:continuity2} to relate $T_+$ to the temperature in front of the shock wave, $T_n$. The first step is then to choose a trial value for $\xi_w$ and solve Eqs.~\eqref{eq:matchingEqs} for $\alpha_+$ and $v_+$. These values can be used as initial conditions\footnote{The initial conditions are given by $v(\xi_w)=\frac{\xi_w-v_+}{1-\xi_w v_+}$ and $\omega(\xi_w)=\omega_n\frac{(1-3\alpha_n)\mu-\nu}{(1-3\alpha_+)\mu-\nu}$. Here $\omega_n$ is a constant and can be conveniently set to one in the numerical calculation. Both the wall velocity and kinetic energy fraction do not depend on it}. to integrate Eqs.~\eqref{eq:continuity2}. We stop the integration when $\xi\frac{\xi-v(\xi)}{1-\xi v(\xi)}=c_s^2$, which is satisfied at the shock front. At first, it will be impossible to satisfy Eq.~\eqref{eq:matchsw} exactly, as the initial trial value of $\xi_w$ may not be appropriate. One can then repeat this procedure with different values of $\xi_w$ until Eq.~\eqref{eq:matchsw} is satisfied at the shock wave.

This algorithm is implemented in the code snippet from lines 5 to 78. The wall velocity can be calculated with the function \texttt{find\_vw}, which uses the function \texttt{root\_scalar} from the library Scipy to determine the optimal $\xi_w$ to solve Eq.~\eqref{eq:matchsw} at the shock wave. The residual of this equation is computed by \texttt{shooting}, which relies on \texttt{solve\_alpha}, \texttt{get\_vp} and \texttt{w\_from\_alpha} to obtain $\alpha_+$, $v_+$ and $\omega_+$, and on \texttt{integrate\_plasma} to integrate Eqs.~\eqref{eq:continuity2} across the shock wave with the Scipy function \texttt{solve\_ivp}. Furthermore, \texttt{find\_vJ} is used to compute the Jouguet velocity, $\texttt{eqWall}$ returns the residual of Eq.~\eqref{eq:matchingEqsB} and \texttt{dfdv} implements Eqs.~\eqref{eq:continuity2}.

To ensure that a solution can be found, \texttt{find\_vw} proceeds with the calculation only if $\alpha_{n,\min}^{\rm def}<\alpha_n<\alpha_{n,\max}^{\rm hyb}$, where $\alpha_{n,\min}^{\rm def}$ is given in Eq.~\eqref{eq:alphamin}. $\alpha_{n,\max}^{\rm hyb}$ is computed by \texttt{max\_al}, which finds the value of $\alpha_n$ for which $\xi_w=\xi_J$.

The efficiency factor $\kappa$ is calculated by \texttt{find\_kappa}. It first determines $\xi_w$ using \texttt{find\_vw}, and then computes the velocity profile in the plasma with \texttt{integrate\_plasma} and evaluates the integral \eqref{eq:rhofl} using the Scipy function \texttt{simps}. Note that we use units for which $\omega_n=1$ in the code.

For completeness, we also include the function \texttt{detonation}, which finds the wall velocity for detonation walls with the LTE approximation. As argued in Sec.~\ref{sec:tem-results}, we do not expect this type of solution to occur in a realistic situation in LTE as a second deflagration or hybrid solution with smaller $\xi_w$ always exists. Moreover, at high wall velocity, out-of-equilibrium effects become more important which can make the LTE approximation less accurate. This function is not needed anywhere else in the code, and can therefore safely be omitted.

\newpage
\begin{lstlisting}[frame=single,language=Python]
import numpy as np
from scipy.integrate import solve_ivp,simps
from scipy.optimize import root_scalar

def find_vJ(alN,cb2): 
    return np.sqrt(cb2)*(1+np.sqrt(3*alN*(1-cb2+3*cb2*alN)))/(1+3*cb2*alN)

def get_vp(vm,al,cb2,branch=-1):
    disc = vm**4-2*cb2*vm**2*(1-6*al)+cb2**2*(1-12*vm**2*al*(1-3*al))
    return 0.5*(cb2+vm**2+branch*np.sqrt(disc))/(vm+3*cb2*vm*al)

def w_from_alpha(al,alN,nu,mu):
    return (abs((1-3*alN)*mu-nu)+1e-100)/(abs((1-3*al)*mu-nu)+1e-100)

def eqWall(al,alN,vm,nu,mu,psiN,solution=-1):
    vp = get_vp(vm,al,1/(nu-1),solution)
    ga2m,ga2p= 1/(1-vm**2),1/(1-vp**2)
    psi = psiN*w_from_alpha(al,alN,nu,mu)**(nu/mu-1)
    return vp*vm*al/(1-(nu-1)*vp*vm)-(1-3*al-(ga2p/ga2m)**(nu/2)*psi)/(3*nu)

def solve_alpha(vw,alN,cb2,cs2,psiN):
    nu,mu = 1+1/cb2,1+1/cs2
    vm = min(np.sqrt(cb2),vw)
    vp_max = min(cs2/vw,vw)
    al_min = max((vm-vp_max)*(cb2-vm*vp_max)/(3*cb2*vm*(1-vp_max**2)),(mu-nu)/(3*mu))
    al_max = 1/3
    branch = -1
    if eqWall(al_min,alN,vm,nu,mu,psiN)*eqWall(al_max,alN,vm,nu,mu,psiN)>0:
        branch = 1
    sol = root_scalar(eqWall,(alN,vm,nu,mu,psiN,branch),bracket=(al_min,al_max),rtol=1e-10,xtol=1e-10)
    if not sol.converged:
        print("WARNING: desired precision not reached in 'solve_alpha'")
    return sol.root

def dfdv(v,X,cs2):
    xi,w = X
    mu_xiv = (xi-v)/(1-xi*v)
    dxidv = xi*(1-v*xi)*(mu_xiv**2/cs2-1)/(2*v*(1-v**2))
    dwdv = w*(1+1/cs2)*mu_xiv/(1-v**2)
    return [dxidv,dwdv]

def integrate_plasma(v0,vw,w0,c2,shock_wave=True):
    def event(v,X,cs2):
        xi,w = X
        return xi*(xi-v)/(1-xi*v) - cs2
    event.terminal = True
    sol = None
    if shock_wave:
        sol = solve_ivp(dfdv,(v0,1e-20),[vw,w0],events=event,args=(c2,),rtol=1e-10,atol=1e-10)
    else:
        sol = solve_ivp(dfdv,(v0,1e-20),[vw,w0],args=(c2,),rtol=1e-10,atol=1e-10)
    if not sol.success:
        print("WARNING: desired precision not reached in 'integrate_plasma'")
    return sol

def shooting(vw,alN,cb2,cs2,psiN):
    nu,mu = 1+1/cb2,1+1/cs2
    vm = min(np.sqrt(cb2),vw)
    al = solve_alpha(vw, alN, cb2, cs2, psiN)
    vp = get_vp(vm, al, cb2)
    wp = w_from_alpha(al, alN, nu, mu)
    sol = integrate_plasma((vw-vp)/(1-vw*vp), vw, wp, cs2)
    vp_sw = sol.y[0,-1]
    vm_sw = (vp_sw-sol.t[-1])/(1-vp_sw*sol.t[-1])
    wm_sw = sol.y[1,-1]
    return vp_sw/vm_sw - ((mu-1)*wm_sw+1)/((mu-1)+wm_sw)

def find_vw(alN,cb2,cs2,psiN):
    nu,mu = 1+1/cb2,1+1/cs2
    vJ = find_vJ(alN, cb2)
    if alN < (1-psiN)/3 or alN <= (mu-nu)/(3*mu):
        print('alN too small')
        return 0
    if alN > max_al(cb2,cs2,psiN,100) or shooting(vJ,alN,cb2,cs2,psiN) < 0:
        print('alN too large')
        return 1
    sol = root_scalar(shooting,(alN,cb2,cs2,psiN),bracket=[1e-3,vJ],rtol=1e-10,xtol=1e-10)
    return sol.root

def max_al(cb2,cs2,psiN,upper_limit=1):
    nu,mu = 1+1/cb2,1+1/cs2
    vm = np.sqrt(cb2)
    def func(alN):
        vw = find_vJ(alN, cb2)
        vp = cs2/vw
        ga2p,ga2m = 1/(1-vp**2),1/(1-vm**2)
        wp = (vp+vw-vw*mu)/(vp+vw-vp*mu)
        psi = psiN*wp**(nu/mu-1)
        al = (mu-nu)/(3*mu)+(alN-(mu-nu)/(3*mu))/wp
        return vp*vm*al/(1-(nu-1)*vp*vm)-(1-3*al-(ga2p/ga2m)**(nu/2)*psi)/(3*nu)
    if func(upper_limit) < 0:
        return upper_limit
    sol = root_scalar(func,bracket=((1-psiN)/3,upper_limit),rtol=1e-10,xtol=1e-10)
    return sol.root

def detonation(alN,cb2,psiN):
    nu = 1+1/cb2
    vJ = find_vJ(alN, cb2)
    def matching_eq(vw):
        A = vw**2+cb2*(1-3*alN*(1-vw**2))
        vm = (A+np.sqrt(A**2-4*vw**2*cb2))/(2*vw)
        ga2w,ga2m = 1/(1-vw**2),1/(1-vm**2)
        return vw*vm*alN/(1-(nu-1)*vw*vm)-(1-3*alN-(ga2w/ga2m)**(nu/2)*psiN)/(3*nu)
    if matching_eq(vJ+1e-10)*matching_eq(1-1e-10) > 0:
        print('No detonation solution')
        return 0
    sol = root_scalar(matching_eq,bracket=(vJ+1e-10,1-1e-10),rtol=1e-10,xtol=1e-10)
    return sol.root

def find_kappa(alN,cb2,cs2,psiN,vw=None):
    if vw is None:
        vw = find_vw(alN,cb2,cs2,psiN)
    nu,mu = 1+1/cb2,1+1/cs2
    kappa,wp,vm,vp = 0,1,0,0
    if vw < 1:
        vm = min(np.sqrt(cb2),vw)
        al = solve_alpha(vw,alN,cb2,cs2,psiN)
        vp = get_vp(vm,al,cb2)
        wp = w_from_alpha(al,alN,nu,mu)
        sol = integrate_plasma((vw-vp)/(1-vw*vp),vw,wp,cs2)
        v,xi,w = sol.t,sol.y[0],sol.y[1]
        kappa += 4*simps((xi*v)**2*w/(1-v**2),xi)/(vw**3*alN)
    if vw**2 > cb2:
        w0 = psiN*wp**(nu/mu)*((1-vm**2)/(1-vp**2))**(nu/2) if vw < 1 else 1+6*alN/(nu-2)
        v0 = (vw-vm)/(1-vw*vm) if vw < 1 else 3*alN/(nu-2+3*alN)
        sol = integrate_plasma(v0,vw,w0,cb2,False)
        v,xi,w = np.flip(sol.t),np.flip(sol.y[0]),np.flip(sol.y[1])
        mask = np.append(xi[1:]>xi[:-1],True)
        kappa += 4*simps(((xi*v)**2*w/(1-v**2))[mask],xi[mask])/(vw**3*alN)
    return kappa
\end{lstlisting}

\end{appendix}

\newpage
\bibliographystyle{utphys}
\bibliography{ref}{}

\providecommand{\href}[2]{#2}\begingroup\raggedright\begin{thebibliography}{10}

\bibitem{Kuzmin:1985mm}
V.~A. Kuzmin, V.~A. Rubakov, and M.~E. Shaposhnikov, ``{On the Anomalous
  Electroweak Baryon Number Nonconservation in the Early Universe},''
  \href{http://dx.doi.org/10.1016/0370-2693(85)91028-7}{{\em Phys. Lett. B}
  {\bfseries 155} (1985) 36}.

\bibitem{Morrissey:2012db}
D.~E. Morrissey and M.~J. Ramsey-Musolf, ``{Electroweak baryogenesis},''
  \href{http://dx.doi.org/10.1088/1367-2630/14/12/125003}{{\em New J. Phys.}
  {\bfseries 14} (2012) 125003},
  \href{http://arxiv.org/abs/1206.2942}{{\ttfamily arXiv:1206.2942 [hep-ph]}}.

\bibitem{Garbrecht:2018mrp}
B.~Garbrecht, ``{Why is there more matter than antimatter? Calculational
  methods for leptogenesis and electroweak baryogenesis},''
  \href{http://dx.doi.org/10.1016/j.ppnp.2019.103727}{{\em Prog. Part. Nucl.
  Phys.} {\bfseries 110} (2020) 103727},
  \href{http://arxiv.org/abs/1812.02651}{{\ttfamily arXiv:1812.02651
  [hep-ph]}}.

\bibitem{Witten:1984rs}
E.~Witten, ``{Cosmic Separation of Phases},''
  \href{http://dx.doi.org/10.1103/PhysRevD.30.272}{{\em Phys. Rev. D}
  {\bfseries 30} (1984) 272--285}.

\bibitem{Kosowsky:1991ua}
A.~Kosowsky, M.~S. Turner, and R.~Watkins, ``{Gravitational radiation from
  colliding vacuum bubbles},''
  \href{http://dx.doi.org/10.1103/PhysRevD.45.4514}{{\em Phys. Rev. D}
  {\bfseries 45} (1992) 4514--4535}.

\bibitem{Kosowsky:1992vn}
A.~Kosowsky and M.~S. Turner, ``{Gravitational radiation from colliding vacuum
  bubbles: envelope approximation to many bubble collisions},''
  \href{http://dx.doi.org/10.1103/PhysRevD.47.4372}{{\em Phys. Rev. D}
  {\bfseries 47} (1993) 4372--4391},
  \href{http://arxiv.org/abs/astro-ph/9211004}{{\ttfamily
  arXiv:astro-ph/9211004}}.

\bibitem{Kamionkowski:1993fg}
M.~Kamionkowski, A.~Kosowsky, and M.~S. Turner, ``{Gravitational radiation from
  first order phase transitions},''
  \href{http://dx.doi.org/10.1103/PhysRevD.49.2837}{{\em Phys. Rev. D}
  {\bfseries 49} (1994) 2837--2851},
  \href{http://arxiv.org/abs/astro-ph/9310044}{{\ttfamily
  arXiv:astro-ph/9310044}}.

\bibitem{Huber:2008hg}
S.~J. Huber and T.~Konstandin, ``{Gravitational Wave Production by Collisions:
  More Bubbles},'' \href{http://dx.doi.org/10.1088/1475-7516/2008/09/022}{{\em
  JCAP} {\bfseries 09} (2008) 022},
  \href{http://arxiv.org/abs/0806.1828}{{\ttfamily arXiv:0806.1828 [hep-ph]}}.

\bibitem{Hindmarsh:2013xza}
M.~Hindmarsh, S.~J. Huber, K.~Rummukainen, and D.~J. Weir, ``{Gravitational
  waves from the sound of a first order phase transition},''
  \href{http://dx.doi.org/10.1103/PhysRevLett.112.041301}{{\em Phys. Rev.
  Lett.} {\bfseries 112} (2014) 041301},
  \href{http://arxiv.org/abs/1304.2433}{{\ttfamily arXiv:1304.2433 [hep-ph]}}.

\bibitem{Grojean:2006bp}
C.~Grojean and G.~Servant, ``{Gravitational Waves from Phase Transitions at the
  Electroweak Scale and Beyond},''
  \href{http://dx.doi.org/10.1103/PhysRevD.75.043507}{{\em Phys. Rev. D}
  {\bfseries 75} (2007) 043507},
  \href{http://arxiv.org/abs/hep-ph/0607107}{{\ttfamily arXiv:hep-ph/0607107}}.

\bibitem{Caprini:2018mtu}
C.~Caprini and D.~G. Figueroa, ``{Cosmological Backgrounds of Gravitational
  Waves},'' \href{http://dx.doi.org/10.1088/1361-6382/aac608}{{\em Class.
  Quant. Grav.} {\bfseries 35} no.~16, (2018) 163001},
  \href{http://arxiv.org/abs/1801.04268}{{\ttfamily arXiv:1801.04268
  [astro-ph.CO]}}.

\bibitem{Caprini:2019egz}
C.~Caprini {\em et~al.}, ``{Detecting gravitational waves from cosmological
  phase transitions with LISA: an update},''
  \href{http://dx.doi.org/10.1088/1475-7516/2020/03/024}{{\em JCAP} {\bfseries
  03} (2020) 024}, \href{http://arxiv.org/abs/1910.13125}{{\ttfamily
  arXiv:1910.13125 [astro-ph.CO]}}.

\bibitem{LISACosmologyWorkingGroup:2022jok}
{\bfseries LISA Cosmology Working Group} Collaboration, P.~Auclair {\em
  et~al.}, ``{Cosmology with the Laser Interferometer Space Antenna},''
  \href{http://arxiv.org/abs/2204.05434}{{\ttfamily arXiv:2204.05434
  [astro-ph.CO]}}.

\bibitem{Cline:2020jre}
J.~M. Cline and K.~Kainulainen, ``{Electroweak baryogenesis at high bubble wall
  velocities},'' \href{http://dx.doi.org/10.1103/PhysRevD.101.063525}{{\em
  Phys. Rev. D} {\bfseries 101} no.~6, (2020) 063525},
  \href{http://arxiv.org/abs/2001.00568}{{\ttfamily arXiv:2001.00568
  [hep-ph]}}.

\bibitem{Cline:2021dkf}
J.~M. Cline and B.~Laurent, ``{Electroweak baryogenesis from light fermion
  sources: A critical study},''
  \href{http://dx.doi.org/10.1103/PhysRevD.104.083507}{{\em Phys. Rev. D}
  {\bfseries 104} no.~8, (2021) 083507},
  \href{http://arxiv.org/abs/2108.04249}{{\ttfamily arXiv:2108.04249
  [hep-ph]}}.

\bibitem{Ellis:2022lft}
J.~Ellis, M.~Lewicki, M.~Merchand, J.~M. No, and M.~Zych, ``{The scalar singlet
  extension of the Standard Model: gravitational waves versus baryogenesis},''
  \href{http://dx.doi.org/10.1007/JHEP01(2023)093}{{\em JHEP} {\bfseries 01}
  (2023) 093}, \href{http://arxiv.org/abs/2210.16305}{{\ttfamily
  arXiv:2210.16305 [hep-ph]}}.

\bibitem{Espinosa:2010hh}
J.~R. Espinosa, T.~Konstandin, J.~M. No, and G.~Servant, ``{Energy Budget of
  Cosmological First-order Phase Transitions},''
  \href{http://dx.doi.org/10.1088/1475-7516/2010/06/028}{{\em JCAP} {\bfseries
  06} (2010) 028}, \href{http://arxiv.org/abs/1004.4187}{{\ttfamily
  arXiv:1004.4187 [hep-ph]}}.

\bibitem{Caprini:2015zlo}
C.~Caprini {\em et~al.}, ``{Science with the space-based interferometer eLISA.
  II: Gravitational waves from cosmological phase transitions},''
  \href{http://dx.doi.org/10.1088/1475-7516/2016/04/001}{{\em JCAP} {\bfseries
  04} (2016) 001}, \href{http://arxiv.org/abs/1512.06239}{{\ttfamily
  arXiv:1512.06239 [astro-ph.CO]}}.

\bibitem{Gowling:2021gcy}
C.~Gowling and M.~Hindmarsh, ``{Observational prospects for phase transitions
  at LISA: Fisher matrix analysis},''
  \href{http://dx.doi.org/10.1088/1475-7516/2021/10/039}{{\em JCAP} {\bfseries
  10} (2021) 039}, \href{http://arxiv.org/abs/2106.05984}{{\ttfamily
  arXiv:2106.05984 [astro-ph.CO]}}.

\bibitem{Steinhardt:1981ct}
P.~J. Steinhardt, ``{Relativistic Detonation Waves and Bubble Growth in False
  Vacuum Decay},'' \href{http://dx.doi.org/10.1103/PhysRevD.25.2074}{{\em Phys.
  Rev. D} {\bfseries 25} (1982) 2074}.

\bibitem{Laine:1993ey}
M.~Laine, ``{Bubble growth as a detonation},''
  \href{http://dx.doi.org/10.1103/PhysRevD.49.3847}{{\em Phys. Rev. D}
  {\bfseries 49} (1994) 3847--3853},
  \href{http://arxiv.org/abs/hep-ph/9309242}{{\ttfamily arXiv:hep-ph/9309242}}.

\bibitem{Kurki-Suonio:1995rrv}
H.~Kurki-Suonio and M.~Laine, ``{Supersonic deflagrations in cosmological phase
  transitions},'' \href{http://dx.doi.org/10.1103/PhysRevD.51.5431}{{\em Phys.
  Rev. D} {\bfseries 51} (1995) 5431--5437},
  \href{http://arxiv.org/abs/hep-ph/9501216}{{\ttfamily arXiv:hep-ph/9501216}}.

\bibitem{Ignatius:1993qn}
J.~Ignatius, K.~Kajantie, H.~Kurki-Suonio, and M.~Laine, ``{The growth of
  bubbles in cosmological phase transitions},''
  \href{http://dx.doi.org/10.1103/PhysRevD.49.3854}{{\em Phys. Rev. D}
  {\bfseries 49} (1994) 3854--3868},
  \href{http://arxiv.org/abs/astro-ph/9309059}{{\ttfamily
  arXiv:astro-ph/9309059}}.

\bibitem{Heckler:1994uu}
A.~F. Heckler, ``{The Effects of electroweak phase transition dynamics on
  baryogenesis and primordial nucleosynthesis},''
  \href{http://dx.doi.org/10.1103/PhysRevD.51.405}{{\em Phys. Rev. D}
  {\bfseries 51} (1995) 405--428},
  \href{http://arxiv.org/abs/astro-ph/9407064}{{\ttfamily
  arXiv:astro-ph/9407064}}.

\bibitem{Kurki-Suonio:1996gkq}
H.~Kurki-Suonio and M.~Laine, ``{Real time history of the cosmological
  electroweak phase transition},''
  \href{http://dx.doi.org/10.1103/PhysRevLett.77.3951}{{\em Phys. Rev. Lett.}
  {\bfseries 77} (1996) 3951--3954},
  \href{http://arxiv.org/abs/hep-ph/9607382}{{\ttfamily arXiv:hep-ph/9607382}}.

\bibitem{Huber:2011aa}
S.~J. Huber and M.~Sopena, ``{The bubble wall velocity in the minimal
  supersymmetric light stop scenario},''
  \href{http://dx.doi.org/10.1103/PhysRevD.85.103507}{{\em Phys. Rev. D}
  {\bfseries 85} (2012) 103507},
  \href{http://arxiv.org/abs/1112.1888}{{\ttfamily arXiv:1112.1888 [hep-ph]}}.

\bibitem{Huber:2013kj}
S.~J. Huber and M.~Sopena, ``{An efficient approach to electroweak bubble
  velocities},'' \href{http://arxiv.org/abs/1302.1044}{{\ttfamily
  arXiv:1302.1044 [hep-ph]}}.

\bibitem{Dine:1992wr}
M.~Dine, R.~G. Leigh, P.~Y. Huet, A.~D. Linde, and D.~A. Linde, ``{Towards the
  theory of the electroweak phase transition},''
  \href{http://dx.doi.org/10.1103/PhysRevD.46.550}{{\em Phys. Rev. D}
  {\bfseries 46} (1992) 550--571},
  \href{http://arxiv.org/abs/hep-ph/9203203}{{\ttfamily arXiv:hep-ph/9203203}}.

\bibitem{Liu:1992tn}
B.-H. Liu, L.~D. McLerran, and N.~Turok, ``{Bubble nucleation and growth at a
  baryon number producing electroweak phase transition},''
  \href{http://dx.doi.org/10.1103/PhysRevD.46.2668}{{\em Phys. Rev. D}
  {\bfseries 46} (1992) 2668--2688}.

\bibitem{Moore:2000wx}
G.~D. Moore, ``{Electroweak bubble wall friction: Analytic results},''
  \href{http://dx.doi.org/10.1088/1126-6708/2000/03/006}{{\em JHEP} {\bfseries
  03} (2000) 006}, \href{http://arxiv.org/abs/hep-ph/0001274}{{\ttfamily
  arXiv:hep-ph/0001274}}.

\bibitem{Moore:1995si}
G.~D. Moore and T.~Prokopec, ``{How fast can the wall move? A Study of the
  electroweak phase transition dynamics},''
  \href{http://dx.doi.org/10.1103/PhysRevD.52.7182}{{\em Phys. Rev. D}
  {\bfseries 52} (1995) 7182--7204},
  \href{http://arxiv.org/abs/hep-ph/9506475}{{\ttfamily arXiv:hep-ph/9506475}}.

\bibitem{Moore:1995ua}
G.~D. Moore and T.~Prokopec, ``{Bubble wall velocity in a first order
  electroweak phase transition},''
  \href{http://dx.doi.org/10.1103/PhysRevLett.75.777}{{\em Phys. Rev. Lett.}
  {\bfseries 75} (1995) 777--780},
  \href{http://arxiv.org/abs/hep-ph/9503296}{{\ttfamily arXiv:hep-ph/9503296}}.

\bibitem{Laurent:2020gpg}
B.~Laurent and J.~M. Cline, ``{Fluid equations for fast-moving electroweak
  bubble walls},'' \href{http://dx.doi.org/10.1103/PhysRevD.102.063516}{{\em
  Phys. Rev. D} {\bfseries 102} no.~6, (2020) 063516},
  \href{http://arxiv.org/abs/2007.10935}{{\ttfamily arXiv:2007.10935
  [hep-ph]}}.

\bibitem{Laurent:2022jrs}
B.~Laurent and J.~M. Cline, ``{First principles determination of bubble wall
  velocity},'' \href{http://dx.doi.org/10.1103/PhysRevD.106.023501}{{\em Phys.
  Rev. D} {\bfseries 106} no.~2, (2022) 023501},
  \href{http://arxiv.org/abs/2204.13120}{{\ttfamily arXiv:2204.13120
  [hep-ph]}}.

\bibitem{Dorsch:2018pat}
G.~C. Dorsch, S.~J. Huber, and T.~Konstandin, ``{Bubble wall velocities in the
  Standard Model and beyond},''
  \href{http://dx.doi.org/10.1088/1475-7516/2018/12/034}{{\em JCAP} {\bfseries
  12} (2018) 034}, \href{http://arxiv.org/abs/1809.04907}{{\ttfamily
  arXiv:1809.04907 [hep-ph]}}.

\bibitem{Wang:2020zlf}
X.~Wang, F.~P. Huang, and X.~Zhang, ``{Bubble wall velocity beyond leading-log
  approximation in electroweak phase transition},''
  \href{http://arxiv.org/abs/2011.12903}{{\ttfamily arXiv:2011.12903
  [hep-ph]}}.

\bibitem{Jiang:2022btc}
S.~Jiang, F.~P. Huang, and X.~Wang, ``{Bubble wall velocity during electroweak
  phase transition in the inert doublet model},''
  \href{http://arxiv.org/abs/2211.13142}{{\ttfamily arXiv:2211.13142
  [hep-ph]}}.

\bibitem{Hoche:2020ysm}
S.~H\"oche, J.~Kozaczuk, A.~J. Long, J.~Turner, and Y.~Wang, ``{Towards an
  all-orders calculation of the electroweak bubble wall velocity},''
  \href{http://dx.doi.org/10.1088/1475-7516/2021/03/009}{{\em JCAP} {\bfseries
  03} (2021) 009}, \href{http://arxiv.org/abs/2007.10343}{{\ttfamily
  arXiv:2007.10343 [hep-ph]}}.

\bibitem{Friedlander:2020tnq}
A.~Friedlander, I.~Banta, J.~M. Cline, and D.~Tucker-Smith, ``{Wall speed and
  shape in singlet-assisted strong electroweak phase transitions},''
  \href{http://dx.doi.org/10.1103/PhysRevD.103.055020}{{\em Phys. Rev. D}
  {\bfseries 103} no.~5, (2021) 055020},
  \href{http://arxiv.org/abs/2009.14295}{{\ttfamily arXiv:2009.14295
  [hep-ph]}}.

\bibitem{Azatov:2020ufh}
A.~Azatov and M.~Vanvlasselaer, ``{Bubble wall velocity: heavy physics
  effects},'' \href{http://dx.doi.org/10.1088/1475-7516/2021/01/058}{{\em JCAP}
  {\bfseries 01} (2021) 058}, \href{http://arxiv.org/abs/2010.02590}{{\ttfamily
  arXiv:2010.02590 [hep-ph]}}.

\bibitem{Cai:2020djd}
R.-G. Cai and S.-J. Wang, ``{Effective picture of bubble expansion},''
  \href{http://dx.doi.org/10.1088/1475-7516/2021/03/096}{{\em JCAP} {\bfseries
  03} (2021) 096}, \href{http://arxiv.org/abs/2011.11451}{{\ttfamily
  arXiv:2011.11451 [astro-ph.CO]}}.

\bibitem{Cline:2021iff}
J.~M. Cline, A.~Friedlander, D.-M. He, K.~Kainulainen, B.~Laurent, and
  D.~Tucker-Smith, ``{Baryogenesis and gravity waves from a UV-completed
  electroweak phase transition},''
  \href{http://dx.doi.org/10.1103/PhysRevD.103.123529}{{\em Phys. Rev. D}
  {\bfseries 103} no.~12, (2021) 123529},
  \href{http://arxiv.org/abs/2102.12490}{{\ttfamily arXiv:2102.12490
  [hep-ph]}}.

\bibitem{Bea:2021zsu}
Y.~Bea, J.~Casalderrey-Solana, T.~Giannakopoulos, D.~Mateos,
  M.~Sanchez-Garitaonandia, and M.~Zilh\~ao, ``{Bubble wall velocity from
  holography},'' \href{http://dx.doi.org/10.1103/PhysRevD.104.L121903}{{\em
  Phys. Rev. D} {\bfseries 104} no.~12, (2021) L121903},
  \href{http://arxiv.org/abs/2104.05708}{{\ttfamily arXiv:2104.05708
  [hep-th]}}.

\bibitem{Bigazzi:2021ucw}
F.~Bigazzi, A.~Caddeo, T.~Canneti, and A.~L. Cotrone, ``{Bubble wall velocity
  at strong coupling},'' \href{http://dx.doi.org/10.1007/JHEP08(2021)090}{{\em
  JHEP} {\bfseries 08} (2021) 090},
  \href{http://arxiv.org/abs/2104.12817}{{\ttfamily arXiv:2104.12817
  [hep-ph]}}.

\bibitem{Lewicki:2021pgr}
M.~Lewicki, M.~Merchand, and M.~Zych, ``{Electroweak bubble wall expansion:
  gravitational waves and baryogenesis in Standard Model-like thermal
  plasma},'' \href{http://dx.doi.org/10.1007/JHEP02(2022)017}{{\em JHEP}
  {\bfseries 02} (2022) 017}, \href{http://arxiv.org/abs/2111.02393}{{\ttfamily
  arXiv:2111.02393 [astro-ph.CO]}}.

\bibitem{Gouttenoire:2021kjv}
Y.~Gouttenoire, R.~Jinno, and F.~Sala, ``{Friction pressure on relativistic
  bubble walls},'' \href{http://dx.doi.org/10.1007/JHEP05(2022)004}{{\em JHEP}
  {\bfseries 05} (2022) 004}, \href{http://arxiv.org/abs/2112.07686}{{\ttfamily
  arXiv:2112.07686 [hep-ph]}}.

\bibitem{Dorsch:2021nje}
G.~C. Dorsch, S.~J. Huber, and T.~Konstandin, ``{A sonic boom in bubble wall
  friction},'' \href{http://dx.doi.org/10.1088/1475-7516/2022/04/010}{{\em
  JCAP} {\bfseries 04} no.~04, (2022) 010},
  \href{http://arxiv.org/abs/2112.12548}{{\ttfamily arXiv:2112.12548
  [hep-ph]}}.

\bibitem{DeCurtis:2022hlx}
S.~De~Curtis, L.~D. Rose, A.~Guiggiani, A.~G. Muyor, and G.~Panico, ``{Bubble
  wall dynamics at the electroweak phase transition},''
  \href{http://dx.doi.org/10.1007/JHEP03(2022)163}{{\em JHEP} {\bfseries 03}
  (2022) 163}, \href{http://arxiv.org/abs/2201.08220}{{\ttfamily
  arXiv:2201.08220 [hep-ph]}}.

\bibitem{Wang:2022txy}
S.-J. Wang and Z.-Y. Yuwen, ``{Hydrodynamic backreaction force of cosmological
  bubble expansion},''
  \href{http://dx.doi.org/10.1103/PhysRevD.107.023501}{{\em Phys. Rev. D}
  {\bfseries 107} no.~2, (2023) 023501},
  \href{http://arxiv.org/abs/2205.02492}{{\ttfamily arXiv:2205.02492
  [hep-ph]}}.

\bibitem{Lewicki:2022nba}
M.~Lewicki, V.~Vaskonen, and H.~Veerm\"ae, ``{Bubble dynamics in fluids with
  N-body simulations},''
  \href{http://dx.doi.org/10.1103/PhysRevD.106.103501}{{\em Phys. Rev. D}
  {\bfseries 106} no.~10, (2022) 103501},
  \href{http://arxiv.org/abs/2205.05667}{{\ttfamily arXiv:2205.05667
  [astro-ph.CO]}}.

\bibitem{Ai:2022kqm}
W.-Y. Ai, J.~S. Cruz, B.~Garbrecht, and C.~Tamarit, ``{Instability of bubble
  expansion at zero temperature},''
  \href{http://dx.doi.org/10.1103/PhysRevD.107.036014}{{\em Phys. Rev. D}
  {\bfseries 107} no.~3, (2023) 036014},
  \href{http://arxiv.org/abs/2209.00639}{{\ttfamily arXiv:2209.00639
  [hep-th]}}.

\bibitem{GarciaGarcia:2022yqb}
I.~Garcia~Garcia, G.~Koszegi, and R.~Petrossian-Byrne, ``{Reflections on Bubble
  Walls},'' \href{http://arxiv.org/abs/2212.10572}{{\ttfamily arXiv:2212.10572
  [hep-ph]}}.

\bibitem{LiLi:2023dlc}
L.~Li, S.-J. Wang, and Z.-Y. Yuwen, ``{Bubble expansion at strong coupling},''
  \href{http://arxiv.org/abs/2302.10042}{{\ttfamily arXiv:2302.10042
  [hep-th]}}.

\bibitem{Krajewski:2023clt}
T.~Krajewski, M.~Lewicki, and M.~Zych, ``{Hydrodynamical constraints on bubble
  wall velocity},'' \href{http://arxiv.org/abs/2303.18216}{{\ttfamily
  arXiv:2303.18216 [astro-ph.CO]}}.

\bibitem{Konstandin:2010dm}
T.~Konstandin and J.~M. No, ``{Hydrodynamic obstruction to bubble expansion},''
  \href{http://dx.doi.org/10.1088/1475-7516/2011/02/008}{{\em JCAP} {\bfseries
  02} (2011) 008}, \href{http://arxiv.org/abs/1011.3735}{{\ttfamily
  arXiv:1011.3735 [hep-ph]}}.

\bibitem{BarrosoMancha:2020fay}
M.~Barroso~Mancha, T.~Prokopec, and B.~Swiezewska, ``{Field-theoretic
  derivation of bubble-wall force},''
  \href{http://dx.doi.org/10.1007/JHEP01(2021)070}{{\em JHEP} {\bfseries 01}
  (2021) 070}, \href{http://arxiv.org/abs/2005.10875}{{\ttfamily
  arXiv:2005.10875 [hep-th]}}.

\bibitem{Balaji:2020yrx}
S.~Balaji, M.~Spannowsky, and C.~Tamarit, ``{Cosmological bubble friction in
  local equilibrium},''
  \href{http://dx.doi.org/10.1088/1475-7516/2021/03/051}{{\em JCAP} {\bfseries
  03} (2021) 051}, \href{http://arxiv.org/abs/2010.08013}{{\ttfamily
  arXiv:2010.08013 [hep-ph]}}.

\bibitem{Ai:2021kak}
W.-Y. Ai, B.~Garbrecht, and C.~Tamarit, ``{Bubble wall velocities in local
  equilibrium},'' \href{http://dx.doi.org/10.1088/1475-7516/2022/03/015}{{\em
  JCAP} {\bfseries 03} no.~03, (2022) 015},
  \href{http://arxiv.org/abs/2109.13710}{{\ttfamily arXiv:2109.13710
  [hep-ph]}}.

\bibitem{landau1987fluid}
L.~D. Landau and E.~M. Lifshitz, {\em Fluid Mechanics}.
\newblock Pergamon Press, New York, 1989.

\bibitem{Giese:2020rtr}
F.~Giese, T.~Konstandin, and J.~van~de Vis, ``{Model-independent energy budget
  of cosmological first-order phase transitions\textemdash{}A sound argument to
  go beyond the bag model},''
  \href{http://dx.doi.org/10.1088/1475-7516/2020/07/057}{{\em JCAP} {\bfseries
  07} no.~07, (2020) 057}, \href{http://arxiv.org/abs/2004.06995}{{\ttfamily
  arXiv:2004.06995 [astro-ph.CO]}}.

\bibitem{Giese:2020znk}
F.~Giese, T.~Konstandin, K.~Schmitz, and J.~van~de Vis, ``{Model-independent
  energy budget for LISA},''
  \href{http://dx.doi.org/10.1088/1475-7516/2021/01/072}{{\em JCAP} {\bfseries
  01} (2021) 072}, \href{http://arxiv.org/abs/2010.09744}{{\ttfamily
  arXiv:2010.09744 [astro-ph.CO]}}.

\bibitem{Leitao:2014pda}
L.~Leitao and A.~Megevand, ``{Hydrodynamics of phase transition fronts and the
  speed of sound in the plasma},''
  \href{http://dx.doi.org/10.1016/j.nuclphysb.2014.12.008}{{\em Nucl. Phys. B}
  {\bfseries 891} (2015) 159--199},
  \href{http://arxiv.org/abs/1410.3875}{{\ttfamily arXiv:1410.3875 [hep-ph]}}.

\bibitem{Tenkanen:2022tly}
T.~V.~I. Tenkanen and J.~van~de Vis, ``{Speed of sound in cosmological phase
  transitions and effect on gravitational waves},''
  \href{http://dx.doi.org/10.1007/JHEP08(2022)302}{{\em JHEP} {\bfseries 08}
  (2022) 302}, \href{http://arxiv.org/abs/2206.01130}{{\ttfamily
  arXiv:2206.01130 [hep-ph]}}.

\bibitem{Laine:2015kra}
M.~Laine and M.~Meyer, ``{Standard Model thermodynamics across the electroweak
  crossover},'' \href{http://dx.doi.org/10.1088/1475-7516/2015/07/035}{{\em
  JCAP} {\bfseries 07} (2015) 035},
  \href{http://arxiv.org/abs/1503.04935}{{\ttfamily arXiv:1503.04935
  [hep-ph]}}.

\bibitem{Ares:2020lbt}
F.~R. Ares, M.~Hindmarsh, C.~Hoyos, and N.~Jokela, ``{Gravitational waves from
  a holographic phase transition},''
  \href{http://dx.doi.org/10.1007/JHEP04(2021)100}{{\em JHEP} {\bfseries 21}
  (2020) 100}, \href{http://arxiv.org/abs/2011.12878}{{\ttfamily
  arXiv:2011.12878 [hep-th]}}.

\bibitem{Janik:2022wsx}
R.~A. Janik, M.~Jarvinen, H.~Soltanpanahi, and J.~Sonnenschein, ``{Perfect
  Fluid Hydrodynamic Picture of Domain Wall Velocities at Strong Coupling},''
  \href{http://dx.doi.org/10.1103/PhysRevLett.129.081601}{{\em Phys. Rev.
  Lett.} {\bfseries 129} no.~8, (2022) 081601},
  \href{http://arxiv.org/abs/2205.06274}{{\ttfamily arXiv:2205.06274
  [hep-th]}}.

\bibitem{Bodeker:2009qy}
D.~Bodeker and G.~D. Moore, ``{Can electroweak bubble walls run away?},''
  \href{http://dx.doi.org/10.1088/1475-7516/2009/05/009}{{\em JCAP} {\bfseries
  05} (2009) 009}, \href{http://arxiv.org/abs/0903.4099}{{\ttfamily
  arXiv:0903.4099 [hep-ph]}}.

\bibitem{Bodeker:2017cim}
D.~Bodeker and G.~D. Moore, ``{Electroweak Bubble Wall Speed Limit},''
  \href{http://dx.doi.org/10.1088/1475-7516/2017/05/025}{{\em JCAP} {\bfseries
  05} (2017) 025}, \href{http://arxiv.org/abs/1703.08215}{{\ttfamily
  arXiv:1703.08215 [hep-ph]}}.

\bibitem{Hindmarsh:2015qta}
M.~Hindmarsh, S.~J. Huber, K.~Rummukainen, and D.~J. Weir, ``{Numerical
  simulations of acoustically generated gravitational waves at a first order
  phase transition},'' \href{http://dx.doi.org/10.1103/PhysRevD.92.123009}{{\em
  Phys. Rev. D} {\bfseries 92} no.~12, (2015) 123009},
  \href{http://arxiv.org/abs/1504.03291}{{\ttfamily arXiv:1504.03291
  [astro-ph.CO]}}.

\bibitem{Hindmarsh:2017gnf}
M.~Hindmarsh, S.~J. Huber, K.~Rummukainen, and D.~J. Weir, ``{Shape of the
  acoustic gravitational wave power spectrum from a first order phase
  transition},'' \href{http://dx.doi.org/10.1103/PhysRevD.96.103520}{{\em Phys.
  Rev. D} {\bfseries 96} no.~10, (2017) 103520},
  \href{http://arxiv.org/abs/1704.05871}{{\ttfamily arXiv:1704.05871
  [astro-ph.CO]}}. [Erratum: Phys.Rev.D 101, 089902 (2020)].

\bibitem{Jinno:2020eqg}
R.~Jinno, T.~Konstandin, and H.~Rubira, ``{A hybrid simulation of gravitational
  wave production in first-order phase transitions},''
  \href{http://dx.doi.org/10.1088/1475-7516/2021/04/014}{{\em JCAP} {\bfseries
  04} (2021) 014}, \href{http://arxiv.org/abs/2010.00971}{{\ttfamily
  arXiv:2010.00971 [astro-ph.CO]}}.

\bibitem{Jinno:2022mie}
R.~Jinno, T.~Konstandin, H.~Rubira, and I.~Stomberg, ``{Higgsless simulations
  of cosmological phase transitions and gravitational waves},''
  \href{http://dx.doi.org/10.1088/1475-7516/2023/02/011}{{\em JCAP} {\bfseries
  02} (2023) 011}, \href{http://arxiv.org/abs/2209.04369}{{\ttfamily
  arXiv:2209.04369 [astro-ph.CO]}}.

\bibitem{Cutting:2019zws}
D.~Cutting, M.~Hindmarsh, and D.~J. Weir, ``{Vorticity, kinetic energy, and
  suppressed gravitational wave production in strong first order phase
  transitions},'' \href{http://dx.doi.org/10.1103/PhysRevLett.125.021302}{{\em
  Phys. Rev. Lett.} {\bfseries 125} no.~2, (2020) 021302},
  \href{http://arxiv.org/abs/1906.00480}{{\ttfamily arXiv:1906.00480
  [hep-ph]}}.

\bibitem{DeCurtis:2023hil}
S.~De~Curtis, L.~Delle~Rose, A.~Guiggiani, A.~Gil~Muyor, and G.~Panico,
  ``{Collision integrals for cosmological phase transitions},''
  \href{http://dx.doi.org/10.1007/JHEP05(2023)194}{{\em JHEP} {\bfseries 05}
  (2023) 194}, \href{http://arxiv.org/abs/2303.05846}{{\ttfamily
  arXiv:2303.05846 [hep-ph]}}.

\end{thebibliography}\endgroup

\end{document}